\documentclass{article}

\input{skilift.sty}

\title{Some improvements to product formula circuits for
  Hamiltonian simulation}
\author{Andre Kornell and Peter Selinger}
\date{Dalhousie University}

\begin{document}
\maketitle

\begin{abstract}
  We provide three improvements to the product formula implementation
  of the ground state energy estimation algorithm via Trotter-Suzuki
  decomposition. These consist of smaller circuit templates for each
  Hamiltonian term, parallelization of commuting controlled rotations,
  and more efficient parallel scheduling. These improvements may be
  regarded separately, and we anticipate that they may be combined
  with other improvements to the product formula implementation.
\end{abstract}

% ----------------------------------------------------------------------
\section{Introduction}

The problem of simulating quantum systems was an initial motivation
for the investigation of quantum computing {\cite{Feynman1982,
    Lloyd1996}}, and it is still regarded to be a likely first
application of quantum computers {\cite{RWSWT2017, CMNRS2018}}. This
computational task is now commonly known as \emph{Hamiltonian
  simulation}. Hamiltonian simulation for individual molecules is
anticipated to speed chemical research in both the academic and the
industrial settings. The quantum phase estimation algorithm can be
used in conjunction with Hamiltonian simulation to estimate the ground
state energy of the molecule. This quantity determines basic chemical
properties such as stability and reactivity. For a general survey
of quantum chemistry simulations on quantum computers, see
{\cite[Section 4]{Cao-et-al-2019}}.

Hamiltonian simulation can be implemented on a universal quantum
computer using methods such as Trotter-Suzuki decomposition
{\cite{BACS2007}} and quantum signal processing {\cite{LC2017}}. There
is ongoing research into the optimization of these methods in specific
settings, e.g., using tensor hypercontraction for quantum chemistry
{\cite{HPM2012, PHMS2012, HPSM2012, LBGHMWB2021}}. One usually tries
to optimize circuits with respect to one or more notions of cost, such
as space, i.e., the number of logical qubits, and time, i.e., the
number of parallel applications of basic gates.

Optimization is specific to the quantum computer architecture, and
there is often a trade-off between space and time in the choice of
architecture and in the choice of optimization. Prominently, the use
of magic states in architectures based on surface codes
{\cite{FMMC2012}} lowers the time-cost of $T$-gates and makes the
time-cost of information transport more significant. Transport of
information from one part of the quantum computer to another is a
ubiquitous feature of circuits in typical quantum computing
architectures because all basic operations are performed locally.

In this paper, we consider three optimizations for those
implementations of molecular Hamiltonian simulation that are based on
Trotter-Suzuki decomposition. The first of these optimizations appears
in the literature, and the other two are original. By combining all
three optimizations, we obtain a highly parallelized circuit.  In
section~\ref{ssec:hamilton}, we provide smaller circuit templates for
each Hamiltonian interaction term by decomposing the term into
rank-two operators rather than into Pauli matrices
{\cite{WLMN2021}}. In
section~\ref{ssec:parallel-controlled-rotations}, we provide a method
to implement controlled $z$-rotations in parallel, which now run in
constant time on surface-code architectures. In
section~\ref{ssec:skilift}, we provide a scheduling method for the
parallelization of Hamiltonian terms. This method could be useful in
the setting of molecules without sparse Hamiltonians.

Bravyi and Kitaev {\cite{BK2002}} described a method for reducing the cost of a
single Hamiltonian term in a molecular simulation from $O(m)$ to
$O(\log m)$, where $m$ is the number of orbitals, essentially by
arranging the orbitals in a tree structure, rather than
linearly. Their method does not immediately allow the circuits for
several Hamiltonian terms to be parallelized. By contrast, our skilift
method keeps the linear arrangement of the orbitals, but uses
fermionic swap gates to efficiently permute the orbitals, so that each
Hamiltonian term can be implemented when its orbital qubits are
adjacent. In this way, we reduce the average gate count per
Hamiltonian term to $O(1)$.  Moreover, our method permits $O(m)$
Hamiltonian terms to be performed in parallel, allowing us to achieve
an average circuit depth per Hamiltonian term of $O(1/m)$.

Note that we are not proposing a new algorithm for ground state energy
estimation, nor are we claiming that the Trotter-Suzuki product
formula family of algorithms is the optimal choice for this
problem. Rather, we are demonstrating the use of circuit optimization
techniques to give a very efficient implementation of this particular
algorithm.

% ----------------------------------------------------------------------
\section{Background}

Ground state energy estimation is the problem of estimating the energy
of the ground state of a molecule whose geometry is assumed to be
known. Thus, the arrangement of the nuclei is given, and the task is
to determine the minimum energy among the possible states of its
electrons. This quantity is significant for the calculation of basic
chemical properties, such as the released energy and the rate of a
reaction. It plays a basic role in industrial research.

The nuclei of the molecule are assumed to be fixed, and the electrons
are assumed to occupy only finitely many low-energy molecular orbitals
$\ket{\psi_0}, \ldots, \ket{\psi_{m-1}}$. The former assumption is
justified by the Born-Oppenheimer approximation and appeals to the
fact that nuclei have far more mass than electrons. The latter
assumption is conceptually justified by the fact that high-energy
orbitals are negligibly occupied in the ground state of the molecule.

The state of the electrons is described by a unit vector in the fermionic Fock space, which has an orthonormal basis consisting of vectors of the form $\ket{\psi_{k_1}} \wedge \cdots \wedge \ket{\psi_{k_p}}$, where $0 \leq k_1 < \cdots < k_p \leq m - 1$. Such a basis vector describes a state in which exactly the orbitals $\ket{\psi_{k_1}}, \ldots, \ket{\psi_{k_p}}$ are occupied. The creation operator $a_k^\dagger$ for orbital $\ket{\psi_k}$ is then defined by $a_k^\dagger \ket{\phi} = \ket{\psi_k} \wedge \ket{\phi}$, and the corresponding annihilation operator $a_k$ is just the adjoint of $a_k^\dagger$.

The Hamiltonian of this system accounts for the attractive Coulomb
force between each nucleus and each electron, the repulsive Coulomb
force between each pair of electrons, and of course the kinetic energy
of electrons. The molecular orbitals are typically chosen to be
real-valued wave functions, and consequently the Hamiltonian is of the
form
\begin{equation*}
H = \sum_{p,q} h_{pq}(a_p^\dagger a_q + a_q^\dagger a_p) + \sum_{p,q,r,s} h_{pqrs} (a_p^\dagger a_q^\dagger a_r a_s + a_s^\dagger a_r^\dagger a_q a_p),
\end{equation*}
where the coefficients $h_{pq}$ and $h_{pqrs}$ are real numbers and the indices $p,q,r,s$ range over $\{0, \ldots, m-1\}$ in a way that makes the terms linearly independent.

The Jordan-Wigner transform yields an implementation of this Fock space on a quantum computer. Formally, it is a unitary operator $\mathcal H \to \mathbb C^2 \otimes \cdots \otimes \mathbb C^2$, where $\mathcal H$ is the Fock space and there are $m$ tensor factors of $\mathbb C^2$.  The Jordan-Wigner transform maps each basis vector $\ket{\psi_{k_1}} \wedge \cdots \wedge \ket{\psi_{k_p}}$ to the basis vector $\ket{b_0} \otimes  \cdots \otimes \ket{b_{m-1}}$, where $b_k=1$ if the orbital $\ket{\psi_k}$ is occupied and otherwise $b_k = 0$. Under this transform, the annihilation operator $a_k$ becomes the matrix $a_k = Z \otimes \cdots \otimes Z \otimes A \otimes I \otimes \cdots \otimes I$, where  $I = (\begin{smallmatrix} 1 & 0 \\ 0 & 1\end{smallmatrix})$, $Z = (\begin{smallmatrix}1 & 0 \\ 0 & -1 \end{smallmatrix})$, and $A = (\begin{smallmatrix} 0 & 1 \\ 0 & 0 \end{smallmatrix})$.

The Jordan-Wigner transform depends nontrivially on the numbering of the molecular orbitals. Let $U$ be the Jordan-Wigner transform for some initial ordering of the orbitals, and let $V$ be the Jordan-Wigner transform after some permutation $\pi$. Then, the unitary matrix $V U^\dagger$ maps each standard basis vector $\ket{b_0} \otimes  \cdots \otimes \ket{b_{m-1}}$ to $\pm\ket{b_{\pi(0)}} \otimes  \cdots \otimes \ket{b_{\pi(m-1)}}$, where the sign depends on the parity of the order permutation of the occupied orbitals.

The time evolution of the electron system is implemented approximately via the fourth-order Trotter-Suzuki decomposition, which is a special case of the Lie product formula. If the Hamiltonian is the sum of two matrices, $H = H_1 + H_2$, then we approximate
\begin{equation*}
e^{-iHt} \approx \prod_{n=1}^N S_2(\alpha/n)S_2(\beta/n)S_2(\alpha/n),
\end{equation*}
where $S_2(x) : = e^{-iH_1tx/2} e^{-iH_2tx} e^{-iH_1tx/2}$ and the real constants $\alpha$ and $\beta$ solve $2\alpha + \beta = 1$ and $2\alpha^3 + \beta^3 = 0$. In our case, the Hamiltonian $H$ is the sum of $\mathcal O (m^4)$ terms, most of which are of the form $h_{pqrs} (a_p^\dagger a_q^\dagger a_r a_s + a_s^\dagger a_r^\dagger a_q a_p)$. The fourth-order Trotter-Suzuki decomposition has a straightforward generalization to this case.

This Hamiltonian simulation is used in combination with the quantum Fourier transform to estimate the ground state energy of the molecule. Intuitively, instead of simulating the evolution of the system for a single time quantity, we simulate the system for a superposition of different time quantities. Thus, we introduce an additional quantum system with some observable $T$, and the composite system evolves as $\ket{\psi} \otimes \ket{T\! =\! t}\mapsto e^{-i H t}\ket{\psi} \otimes \ket{T\! =\! t}$. This is implemented by introducing ``precision qubits,'' which store the binary digits of $t$.

Over the course of the computation, the precision qubits and the orbital qubits become entangled. If the precision qubits are initialized in a superposition of eigenstates of $T$ with equal weight and the orbital qubits are initialized in a state close to the ground state, then the reduced state of the precision qubits is likely to be $2^{-p/2}\sum_{k = 1}^{2^p} e^{-iE_0t_k}\ket{T\!=\!t_k}$, where $E_0$ is the ground state energy and $p$ is the number of precision qubits. Furthermore, if $t_k = kt_1$, then the inverse Fourier transform of the reduced state will peak at $\ket{\check T \! = \! s}$, where $s$ is the binary representation of $E_0$ after the radix point for a unit of energy that depends on~$t_1$.

Thus, implementing the Hamiltonian simulation and the quantum Fourier transform and then measuring precision qubits yields a segment of the binary representation of $E_0$ that is determined by the number of precision qubits $p$ and the smallest evolution time $t_1$. 

\section{Improved circuits for Hamiltonian simulation}

% ----------------------------------------------------------------------
\subsection{The starting point}

The straightforward implementation of this ground state energy estimation algorithm is inefficient in a number of respects. We highlight three of them and explain our improvements.

First, the terms of the Hamiltonian, such as $h_{pqrs} (a_p^\dagger a_q^\dagger a_r a_s + a_s^\dagger a_r^\dagger a_q a_p)$, are traditionally decomposed as linear combinations of Pauli operators {\cite{WBA2011}}. This is because a matrix of the form $e^{i \theta \sigma}$ can be simply implemented for each angle $\theta$ and each Pauli operator $\sigma$. This decomposition increases the number of terms and hence the run time by a factor of eight. We avoid this factor of eight by avoiding this linear decomposition.

Second, in previous implementations of this algorithm, the Trotter-Suzuki decomposition was implemented in series, as they appear in the decomposition. However, many of the factors in the Trotter-Suzuki decomposition commute because they refer to disjoint sets of molecular orbitals. Furthermore, each factor of the decomposition naively corresponds to many controlled rotations when there is more than a single precision qubit, and these controlled rotations also commute. We implement these commuting rotations in a parallel, decreasing the run time by a factor of $O(m)$.

Third, in order to implement a commuting set of factors in the Trotter-Suzuki decomposition in parallel, we reorder the orbital qubits so as to group the qubits that correspond to each factor. It is necessary to reorder the qubits in this way because, while Hamiltonian terms that refer to disjoint sets of orbitals do commute, their natural circuit implementations may overlap as an artifact of the Jordan-Wigner transform. Naively, we would reorder the orbital qubits according to an arbitrary sequence of partitions into singletons, pairs, triples, and quadruples. Instead, we sequence the partitions by combining the circle method for round-robin tournaments with a pairing method suggested by Nazarov and Speyer {\cite{Nazarov-Speyer}}.

% ----------------------------------------------------------------------
\subsection{Better circuit templates}
\label{ssec:hamilton}

In many situations, it is useful for the Hamiltonian to be given as a
linear combination of Pauli operators. This leads to a natural but
inefficient circuit representation. For example, consider the term
\begin{equation}
  H_{0132} = h_{0132}(a\da_0a\da_1a_3a_2 + a\da_2a\da_3a_1a_0).
\end{equation}
After applying the Jordan-Wigner transform, this becomes
\begin{equation}\label{eqn-bct-2}
  H_{0132} = h_{0132}(A\da\x A\da\x A\x A+A\x A\x
  A\da\x A\da).
\end{equation}
Note that $A$ and $A\da$ are not Pauli operators, but can be
decomposed into Paulis as $A=\frac{1}{2}(X+iY)$ and
$A\da=\frac{1}{2}(X-iY)$. Substituting this into {\eqref{eqn-bct-2}}
and simplifying, we obtain the expression
\begin{equation}\label{eqn-bct-3}
  H_{0132} = \frac{1}{8}\,h_{0132}\left(
  \begin{array}{l@{~}l@{~}l@{~}l@{}}
    &X\x X\x X\x X&
    -&X\x X\x Y\x Y\\
    +&X\x Y\x X\x Y&
    +&X\x Y\x Y\x X\\
    +&Y\x X\x X\x Y&
    +&Y\x X\x Y\x X\\
    -&Y\x Y\x X\x X&
    +&Y\x Y\x Y\x Y
  \end{array}
  \right),
\end{equation}
which is a linear combination of eight Paulis. Since the eight Pauli
operators commute, the matrix exponential $e^{-iH_{0132}t}$ can be
exactly written as a product of eight terms
$e^{-\frac{i}{8}h_{0132}tX\x X\x X\x X} \cdots
e^{-\frac{i}{8}h_{0132}tY\x Y\x Y\x Y}$. Each of these eight factors
can then be easily written as a quantum circuit via suitable basis
changes, resulting in the following circuit:
\begin{equation}\label{eqn-bct-4}
  \m{\scalebox{0.145}{
    \begin{qcircuit}[scale=0.6]
      \leftlabel{$0$}{0,3};
      \leftlabel{$1$}{0,2};
      \leftlabel{$2$}{0,1};
      \leftlabel{$3$}{0,0};
      \grid{125}{0,1,2,3};
      \def\xx{1}
      \gate{$H$}{\xx+0,0};
      \gate{$H$}{\xx+0,1};
      \gate{$H$}{\xx+0,2};
      \gate{$H$}{\xx+0,3};
      \controlledX{0}{\notgate}{\xx+1.5,1};
      \controlledX{1}{\notgate}{\xx+2.5,2};
      \controlledX{2}{\notgate}{\xx+3.5,3};
      \widegate{$e^{-i\frac{\theta}{8}Z}$}{1}{\xx+5.5,3};
      \controlledX{2}{\notgate}{\xx+7.5,3};
      \controlledX{1}{\notgate}{\xx+8.5,2};
      \controlledX{0}{\notgate}{\xx+9.5,1};
      \gate{$H$}{\xx+11,0};
      \gate{$H$}{\xx+11,1};
      \gate{$H$}{\xx+11,2};
      \gate{$H$}{\xx+11,3};
      \def\xx{15.5}
      \gate{$S\da$}{\xx-1.5,1};
      \gate{$S\da$}{\xx-1.5,0};
      \gate{$H$}{\xx+0,0};
      \gate{$H$}{\xx+0,1};
      \gate{$H$}{\xx+0,2};
      \gate{$H$}{\xx+0,3};
      \controlledX{0}{\notgate}{\xx+1.5,1};
      \controlledX{1}{\notgate}{\xx+2.5,2};
      \controlledX{2}{\notgate}{\xx+3.5,3};
      \widegate{$e^{+i\frac{\theta}{8}Z}$}{1}{\xx+5.5,3};
      \controlledX{2}{\notgate}{\xx+7.5,3};
      \controlledX{1}{\notgate}{\xx+8.5,2};
      \controlledX{0}{\notgate}{\xx+9.5,1};
      \gate{$H$}{\xx+11,0};
      \gate{$H$}{\xx+11,1};
      \gate{$H$}{\xx+11,2};
      \gate{$H$}{\xx+11,3};
      \gate{$S$}{\xx+12.5,1};
      \gate{$S$}{\xx+12.5,0};
      \def\xx{31.5}
      \gate{$S\da$}{\xx-1.5,2};
      \gate{$S\da$}{\xx-1.5,0};
      \gate{$H$}{\xx+0,0};
      \gate{$H$}{\xx+0,1};
      \gate{$H$}{\xx+0,2};
      \gate{$H$}{\xx+0,3};
      \controlledX{0}{\notgate}{\xx+1.5,1};
      \controlledX{1}{\notgate}{\xx+2.5,2};
      \controlledX{2}{\notgate}{\xx+3.5,3};
      \widegate{$e^{-i\frac{\theta}{8}Z}$}{1}{\xx+5.5,3};
      \controlledX{2}{\notgate}{\xx+7.5,3};
      \controlledX{1}{\notgate}{\xx+8.5,2};
      \controlledX{0}{\notgate}{\xx+9.5,1};
      \gate{$H$}{\xx+11,0};
      \gate{$H$}{\xx+11,1};
      \gate{$H$}{\xx+11,2};
      \gate{$H$}{\xx+11,3};
      \gate{$S$}{\xx+12.5,2};
      \gate{$S$}{\xx+12.5,0};
      \def\xx{47.5}
      \gate{$S\da$}{\xx-1.5,2};
      \gate{$S\da$}{\xx-1.5,1};
      \gate{$H$}{\xx+0,0};
      \gate{$H$}{\xx+0,1};
      \gate{$H$}{\xx+0,2};
      \gate{$H$}{\xx+0,3};
      \controlledX{0}{\notgate}{\xx+1.5,1};
      \controlledX{1}{\notgate}{\xx+2.5,2};
      \controlledX{2}{\notgate}{\xx+3.5,3};
      \widegate{$e^{-i\frac{\theta}{8}Z}$}{1}{\xx+5.5,3};
      \controlledX{2}{\notgate}{\xx+7.5,3};
      \controlledX{1}{\notgate}{\xx+8.5,2};
      \controlledX{0}{\notgate}{\xx+9.5,1};
      \gate{$H$}{\xx+11,0};
      \gate{$H$}{\xx+11,1};
      \gate{$H$}{\xx+11,2};
      \gate{$H$}{\xx+11,3};
      \gate{$S$}{\xx+12.5,2};
      \gate{$S$}{\xx+12.5,1};
      \def\xx{63.5}
      \gate{$S\da$}{\xx-1.5,3};
      \gate{$S\da$}{\xx-1.5,0};
      \gate{$H$}{\xx+0,0};
      \gate{$H$}{\xx+0,1};
      \gate{$H$}{\xx+0,2};
      \gate{$H$}{\xx+0,3};
      \controlledX{0}{\notgate}{\xx+1.5,1};
      \controlledX{1}{\notgate}{\xx+2.5,2};
      \controlledX{2}{\notgate}{\xx+3.5,3};
      \widegate{$e^{-i\frac{\theta}{8}Z}$}{1}{\xx+5.5,3};
      \controlledX{2}{\notgate}{\xx+7.5,3};
      \controlledX{1}{\notgate}{\xx+8.5,2};
      \controlledX{0}{\notgate}{\xx+9.5,1};
      \gate{$H$}{\xx+11,0};
      \gate{$H$}{\xx+11,1};
      \gate{$H$}{\xx+11,2};
      \gate{$H$}{\xx+11,3};
      \gate{$S$}{\xx+12.5,3};
      \gate{$S$}{\xx+12.5,0};
      \def\xx{79.5}
      \gate{$S\da$}{\xx-1.5,3};
      \gate{$S\da$}{\xx-1.5,1};
      \gate{$H$}{\xx+0,0};
      \gate{$H$}{\xx+0,1};
      \gate{$H$}{\xx+0,2};
      \gate{$H$}{\xx+0,3};
      \controlledX{0}{\notgate}{\xx+1.5,1};
      \controlledX{1}{\notgate}{\xx+2.5,2};
      \controlledX{2}{\notgate}{\xx+3.5,3};
      \widegate{$e^{-i\frac{\theta}{8}Z}$}{1}{\xx+5.5,3};
      \controlledX{2}{\notgate}{\xx+7.5,3};
      \controlledX{1}{\notgate}{\xx+8.5,2};
      \controlledX{0}{\notgate}{\xx+9.5,1};
      \gate{$H$}{\xx+11,0};
      \gate{$H$}{\xx+11,1};
      \gate{$H$}{\xx+11,2};
      \gate{$H$}{\xx+11,3};
      \gate{$S$}{\xx+12.5,3};
      \gate{$S$}{\xx+12.5,1};
      \def\xx{95.5}
      \gate{$S\da$}{\xx-1.5,3};
      \gate{$S\da$}{\xx-1.5,2};
      \gate{$H$}{\xx+0,0};
      \gate{$H$}{\xx+0,1};
      \gate{$H$}{\xx+0,2};
      \gate{$H$}{\xx+0,3};
      \controlledX{0}{\notgate}{\xx+1.5,1};
      \controlledX{1}{\notgate}{\xx+2.5,2};
      \controlledX{2}{\notgate}{\xx+3.5,3};
      \widegate{$e^{+i\frac{\theta}{8}Z}$}{1}{\xx+5.5,3};
      \controlledX{2}{\notgate}{\xx+7.5,3};
      \controlledX{1}{\notgate}{\xx+8.5,2};
      \controlledX{0}{\notgate}{\xx+9.5,1};
      \gate{$H$}{\xx+11,0};
      \gate{$H$}{\xx+11,1};
      \gate{$H$}{\xx+11,2};
      \gate{$H$}{\xx+11,3};
      \gate{$S$}{\xx+12.5,3};
      \gate{$S$}{\xx+12.5,2};
      \def\xx{111.5}
      \gate{$S\da$}{\xx-1.5,3};
      \gate{$S\da$}{\xx-1.5,2};
      \gate{$S\da$}{\xx-1.5,1};
      \gate{$S\da$}{\xx-1.5,0};
      \gate{$H$}{\xx+0,0};
      \gate{$H$}{\xx+0,1};
      \gate{$H$}{\xx+0,2};
      \gate{$H$}{\xx+0,3};
      \controlledX{0}{\notgate}{\xx+1.5,1};
      \controlledX{1}{\notgate}{\xx+2.5,2};
      \controlledX{2}{\notgate}{\xx+3.5,3};
      \widegate{$e^{-i\frac{\theta}{8}Z}$}{1}{\xx+5.5,3};
      \controlledX{2}{\notgate}{\xx+7.5,3};
      \controlledX{1}{\notgate}{\xx+8.5,2};
      \controlledX{0}{\notgate}{\xx+9.5,1};
      \gate{$H$}{\xx+11,0};
      \gate{$H$}{\xx+11,1};
      \gate{$H$}{\xx+11,2};
      \gate{$H$}{\xx+11,3};
      \gate{$S$}{\xx+12.5,3};
      \gate{$S$}{\xx+12.5,2};
      \gate{$S$}{\xx+12.5,1};
      \gate{$S$}{\xx+12.5,0};
    \end{qcircuit}
    }}
\end{equation}
We can find an equivalent circuit that is eight times smaller by
bypassing the Pauli decomposition and working directly from
{\eqref{eqn-bct-2}}. Note that $A=\ket{0}\bra{1}$ and
$A\da=\ket{1}\bra{0}$, and therefore
\begin{equation}\label{eqn-bct-5}
  H_{0132} = h_{0132}(\ket{1100}\bra{0011} + \ket{0011}\bra{1100}).
\end{equation}
The key observation is that the matrix {\eqref{eqn-bct-5}} is of rank
2, and has a much more compact circuit decomposition than
{\eqref{eqn-bct-3}}, which is a sum of eight matrices of rank 16.
Specifically, the operation $e^{-i\theta (\ket{1100}\bra{0011} +
  \ket{0011}\bra{1100})}$ can be represented by the following circuit:
\begin{equation}\label{eqn-bct-6}
  \m{\scalebox{0.8}{
    \begin{qcircuit}[scale=0.6]
      \leftlabel{$0$}{0,3};
      \leftlabel{$1$}{0,2};
      \leftlabel{$2$}{0,1};
      \leftlabel{$3$}{0,0};
      \grid{13}{0,1,2,3};
      \def\xx{1}
      \controlledX{0}{\notgate}{\xx+0,1};
      \controlledX{0}{\notgate}{\xx+1,2};
      \controlledX{0}{\notgate}{\xx+2,3};
      \gate{$H$}{\xx+3.25,0};
      \controlledX{2,3}{\wcontrolledX{1}{\widegate{$e^{-i\theta Z}$}{1}}}{\xx+5.5,0};
      \gate{$H$}{\xx+7.75,0};
      \controlledX{0}{\notgate}{\xx+9,3};
      \controlledX{0}{\notgate}{\xx+10,2};
      \controlledX{0}{\notgate}{\xx+11,1};
    \end{qcircuit}
  }}
\end{equation}
Compared to {\eqref{eqn-bct-4}}, the circuit {\eqref{eqn-bct-6}} is
much smaller. On the other hand, it contains a triply-controlled
rotation, rather than an uncontrolled rotation, which a priori
requires more gates. However, it turns out that the latter makes no
difference: in the context of quantum phase estimation, all of the
rotations need to be controlled anyway, and we have efficient ways of
implementing multiply-controlled rotations (see
Section~\ref{ssec:parallel-controlled-rotations} below). The same
optimization was proposed in {\cite[Fig.~2]{WLMN2021}}.

Another notable feature of {\eqref{eqn-bct-5}} and {\eqref{eqn-bct-6}}
is that the related Hamiltonian terms $H_{0231}$ and $H_{0321}$
can be implemented by almost identical circuits. Indeed, these
correspond, respectively, to the operations
$e^{-i\theta' (\ket{1010}\bra{0101} + \ket{0101}\bra{1010})}$
and
$e^{-i\theta'' (\ket{1001}\bra{0110} + \ket{0110}\bra{1001})}$,
and can be implemented by the following respective circuits:
\begin{equation}\label{eqn-bct-7}
  \m{\scalebox{0.8}{
    \begin{qcircuit}[scale=0.6]
      \leftlabel{$0$}{0,3};
      \leftlabel{$1$}{0,2};
      \leftlabel{$2$}{0,1};
      \leftlabel{$3$}{0,0};
      \grid{13}{0,1,2,3};
      \def\xx{1}
      \controlledX{0}{\notgate}{\xx+0,1};
      \controlledX{0}{\notgate}{\xx+1,2};
      \controlledX{0}{\notgate}{\xx+2,3};
      \gate{$H$}{\xx+3.25,0};
      \controlledX{1,3}{\wcontrolledX{2}{\widegate{$e^{-i\theta' Z}$}{1}}}{\xx+5.5,0};
      \gate{$H$}{\xx+7.75,0};
      \controlledX{0}{\notgate}{\xx+9,3};
      \controlledX{0}{\notgate}{\xx+10,2};
      \controlledX{0}{\notgate}{\xx+11,1};
    \end{qcircuit}
  }}
\end{equation}
\begin{equation}\label{eqn-bct-8}
  \m{\scalebox{0.8}{
    \begin{qcircuit}[scale=0.6]
      \leftlabel{$0$}{0,3};
      \leftlabel{$1$}{0,2};
      \leftlabel{$2$}{0,1};
      \leftlabel{$3$}{0,0};
      \grid{13}{0,1,2,3};
      \def\xx{1}
      \controlledX{0}{\notgate}{\xx+0,1};
      \controlledX{0}{\notgate}{\xx+1,2};
      \controlledX{0}{\notgate}{\xx+2,3};
      \gate{$H$}{\xx+3.25,0};
      \controlledX{1,2}{\wcontrolledX{3}{\widegate{$e^{-i\theta'' Z}$}{1}}}{\xx+5.5,0};
      \gate{$H$}{\xx+7.75,0};
      \controlledX{0}{\notgate}{\xx+9,3};
      \controlledX{0}{\notgate}{\xx+10,2};
      \controlledX{0}{\notgate}{\xx+11,1};
    \end{qcircuit}
  }}
\end{equation}
In fact, all three operations can be implemented with a single common
basis change:
\begin{equation}\label{eqn-bct-9}
  \m{\scalebox{0.8}{
    \begin{qcircuit}[scale=0.6]
      \leftlabel{$0$}{0,3};
      \leftlabel{$1$}{0,2};
      \leftlabel{$2$}{0,1};
      \leftlabel{$3$}{0,0};
      \grid{18}{0,1,2,3};
      \def\xx{1}
      \controlledX{0}{\notgate}{\xx+0,1};
      \controlledX{0}{\notgate}{\xx+1,2};
      \controlledX{0}{\notgate}{\xx+2,3};
      \gate{$H$}{\xx+3.25,0};
      \controlledX{2,3}{\wcontrolledX{1}{\widegate{$e^{-i\theta Z}$}{1}}}{\xx+5.5,0};
      \controlledX{1,3}{\wcontrolledX{2}{\widegate{$e^{-i\theta' Z}$}{1}}}{\xx+8,0};
      \controlledX{1,2}{\wcontrolledX{3}{\widegate{$e^{-i\theta'' Z}$}{1}}}{\xx+10.5,0};
      \gate{$H$}{\xx+12.75,0};
      \controlledX{0}{\notgate}{\xx+14,3};
      \controlledX{0}{\notgate}{\xx+15,2};
      \controlledX{0}{\notgate}{\xx+16,1};
    \end{qcircuit}
  }}
\end{equation}
Moreover, the three controlled rotations in the center are diagonal
operators and can be performed in parallel.

% ----------------------------------------------------------------------
\subsection{Parallel controlled rotations}
\label{ssec:parallel-controlled-rotations}

Consider a number of $z$-rotations that are controlled by various
qubits.  Since all controlled $z$-rotations are diagonal gates in the
computational basis, they all commute with each other, so in
principle, they can all be performed in parallel. Here, we consider
advantageous ways to actually perform them in parallel in a
fault-tolerant regime.

We start with the simplest case of a $z$-rotation controlled by a
single qubit.
\[
\m{\scalebox{0.8}{
    \begin{qcircuit}[scale=0.6]
      \grid{4}{0,1};
      \def\xx{0}
      \controlledX{1}{\widegate{$e^{-i\theta Z}$}{1}}{\xx+2,0};
    \end{qcircuit}
}}
\]
A good way to compile this is to decompose it into two uncontrolled
rotations, as follows:
\[
\mp{0.6}{\scalebox{0.8}{
    \begin{qcircuit}[scale=0.6]
      \grid{4}{0,1};
      \def\xx{0}
      \controlledX{1}{\widegate{$e^{-i\theta Z}$}{1}}{\xx+2,0};
    \end{qcircuit}
}}
\quad=\quad
\m{\scalebox{0.8}{
    \begin{qcircuit}[scale=0.6]
      \grid{6}{0,1};
      \def\xx{0}
      \wcontrolledX{0}{\notgate}{\xx+1,1};
      \widegate{$e^{-i\frac{\theta}{2}Z}$}{1}{\xx+3,0};
      \widegate{$e^{-i\frac{\theta}{2}Z}$}{1}{\xx+3,1};
      \wcontrolledX{0}{\notgate}{\xx+5,1};
    \end{qcircuit}
}}
\]
Each uncontrolled rotation can then be fault-tolerantly implemented,
for example by the Ross-Selinger approximate synthesis algorithm
{\cite{RS2016-gridsynth}} or by a fallback method {\cite{BRS2015}}.

When we have more than one control, we can use Toffoli gates and an
ancilla to first consolidate the multiple controls into a single
one. The Toffoli gates require a number of $T$-gates, but that number
is small compared to the number of $T$-gates required to implement the
rotations themselves.
\[
\m{\scalebox{0.8}{
    \begin{qcircuit}[scale=0.6]
      \grid{4}{1,2,3,4};
      \def\xx{0}
      \controlledX{3,4}{\wcontrolledX{2}{\widegate{$e^{-i\theta Z}$}{1}}}{\xx+2,1};
    \end{qcircuit}
}}
\quad=\quad
\m{\scalebox{0.8}{
    \begin{qcircuit}[scale=0.6]
      \grid{7}{0,2,3,4};
      \gridx{1}{6}{1};
      \init{0}{1,1};
      \term{0}{6,1};
      \def\xx{0}
      \controlledX{3,4}{\wcontrolledX{2}{\notgate}}{\xx+2,1};
      \controlledX{1}{\widegate{$e^{-i\theta Z}$}{1}}{\xx+3.5,0};
      \controlledX{3,4}{\wcontrolledX{2}{\notgate}}{\xx+5,1};
    \end{qcircuit}
}}
\]

When we have several rotations targeted at different qubits, we can
perform them in parallel, even if they share controls:
\[
\m{\scalebox{0.8}{
    \begin{qcircuit}[scale=0.6]
      \grid{6.5}{0,1,2,3};
      \def\xx{0}
      \controlledX{3}{\wcontrolledX{2}{\widegate{$e^{-i\theta Z}$}{1}}}{\xx+2,1};
      \controlledX{2}{\wcontrolledX{1}{\widegate{$e^{-i\theta' Z}$}{1}}}{\xx+4.5,0};
    \end{qcircuit}
}}
\quad=\quad
\m{\scalebox{0.8}{
    \begin{qcircuit}[scale=0.6]
      \grid{10}{0,2,4,5};
      \gridx{1}{9}{1,3};
      \init{0}{1,1};
      \init{0}{1,3};
      \term{0}{9,1};
      \term{0}{9,3};
      \def\xx{0}
      \controlledX{5}{\wcontrolledX{4}{\notgate}}{\xx+2,3};
      \controlledX{4}{\wcontrolledX{2}{\notgate}}{\xx+3,1};
      \controlledX{3}{\widegate{$e^{-i\theta Z}$}{1}}{\xx+5,2};
      \controlledX{1}{\widegate{$e^{-i\theta' Z}$}{1}}{\xx+5,0};
      \controlledX{4}{\wcontrolledX{2}{\notgate}}{\xx+7,1};
      \controlledX{5}{\wcontrolledX{4}{\notgate}}{\xx+8,3};
    \end{qcircuit}
}}
\]

An interesting optimization is possible when we have $n$ rotations
that are targeted at the same qubit, possibly using different rotation
angles, but controlled by different qubits. In this case, we only
need $n+1$ uncontrolled rotations:
\[
\m{\scalebox{0.8}{
    \begin{qcircuit}[scale=0.6]
      \grid{9}{0,1,2,3};
      \def\xx{0}
      \controlledX{1}{\widegate{$e^{-i\theta_1 Z}$}{1}}{\xx+2,0};
      \controlledX{2}{\widegate{$e^{-i\theta_2 Z}$}{1}}{\xx+4.5,0};
      \controlledX{3}{\widegate{$e^{-i\theta_3 Z}$}{1}}{\xx+7,0};
    \end{qcircuit}
}}
\quad=\quad
\m{\scalebox{0.8}{
    \begin{qcircuit}[scale=0.6]
      \grid{11}{0,1,2,3};
      \def\xx{0}
      \wcontrolledX{0}{\notgate}{\xx+1,1};
      \wcontrolledX{0}{\notgate}{\xx+2,2};
      \wcontrolledX{0}{\notgate}{\xx+3,3};
      \widegate{$e^{-i\frac{\theta_3}{2} Z}$}{1}{\xx+5.5,3};
      \widegate{$e^{-i\frac{\theta_2}{2} Z}$}{1}{\xx+5.5,2};
      \widegate{$e^{-i\frac{\theta_1}{2} Z}$}{1}{\xx+5.5,1};
      \widegate{$e^{-i\frac{\theta_1+\theta_2+\theta_3}{2} Z}$}{1.7}{\xx+5.5,0};
      \wcontrolledX{0}{\notgate}{\xx+8,3};
      \wcontrolledX{0}{\notgate}{\xx+9,2};
      \wcontrolledX{0}{\notgate}{\xx+10,1};
    \end{qcircuit}
}}
\]

Combining the above methods allows us to perform any number of
controlled $z$-rotations in parallel. Moreover, in the lattice surgery
setting, fanout can be performed in a single time step, allowing
ancillas to be copied instantaneously. Therefore, we can perform any
number of controlled $z$-rotations in constant time (i.e., the time
depends only on the approximation accuracy $\epsilon$, but not on the
number of parallel rotations).

% ----------------------------------------------------------------------
\subsection{Ski lift parallelization}
\label{ssec:skilift}

The Jordan-Wigner transform maps each annihilation operator $a_p$ to the matrix $Z \tensor \cdots \tensor Z \tensor A \tensor I \tensor \cdots \tensor I$, where the matrix $A$ is in position $p$. Recall that $A = (\begin{smallmatrix} 0 & 1 \\ 0 & 0 \end{smallmatrix})$. Consequently, a Hamiltonian term of the form $h_{pq}(a_p^\dagger a_q + a_q^\dagger a_p)$ for $p < q$ is mapped to a matrix that acts nontrivially on all orbital qubits $r$ for $p < r < q$. Thus, if $p_1 < p_2 < q_1 < q_2$, then the circuits implementing the time evolution for Hamiltonian terms $h_{p_1q_1}(a_{p_1}^\dagger a_{q_1} + a_{q_1}^\dagger a_{p_1})$ and $h_{p_2q_2}(a_{p_2}^\dagger a_{q_2} + a_{q_2}^\dagger a_{p_2})$ overlap. Overlapping circuits cannot be executed in parallel. However, if $p_1 < q_1 < p_2 < q_2$, then the circuits implementing the time evolution for these two Hamiltonian terms do not overlap, and hence they can be executed in parallel. We use fermionic swap operators to reduce the first case to the second.

For molecular orbitals $p \neq q$, we can define the swap operator $s_{p,q}$ on the electronic Fock space by $s_{p,q}\ket{\psi_p} = \ket{\psi_q}$, $s_{p,q}\ket{\psi_q} = \ket{\psi_p}$, and $s_{p,q}\ket{\psi_r} = \ket{\psi_r}$ for $r \neq p, q$. Of course, $s_{p,q} a_p s_{p,q} = a_q$. Thus, we can use these swap operators to implement a change of basis in which the circuits implementing the time evolution for Hamiltonian terms $h_{p_1q_1}(a_{p_1}^\dagger a_{q_1} + a_{q_1}^\dagger a_{p_1})$ and $h_{p_2q_2}(a_{p_2}^\dagger a_{q_2} + a_{q_2}^\dagger a_{p_2})$ do not overlap. If $p_1  < p_2 < q_1 < q_2$, then such a change of basis is clearly achieved by $s_{p_2,q_1}$. The same obstacle occurs for Hamiltonian terms involving three or four distinct qubits, and the same solution applies. We refer to such a change of basis as a fermionic permutation.

The Jordan-Wigner transform of the swap operator $s_{p,q}$ has a
simple form. It is a self-adjoint unitary operator that commutes with
$a_r$ for $r \neq p, q$ and satisfies $s_{p,q} a_p s_{p, q} =
a_q$. These two properties imply that $s_{p,p+1}$ has the following
circuit implementation after the Jordan-Wigner transform:
\[
s_{p,p+1}
\quad = \quad
\m{\begin{qcircuit}[scale = 0.4]
\grid{2}{0,1}
\leftlabel{\footnotesize $p$}{0,1}
\leftlabel{\footnotesize $p+1$}{0,0}
\dotswapgate{1}{0}{1}
\end{qcircuit}}
\qquad = \qquad
\m{\begin{qcircuit}[scale = 0.4]
\grid{3.5}{0,1}
\swapgate{1}{0}{1}
\controlledX{1}{\gate{$Z$}}{2.5,0}
\end{qcircuit}}
\]
Since the fermionic transposition operators $s_{p,p+1}$ satisfy the
braid relations, we can define the more general $s_{p,q}$ in terms of them. For example, $s_{p, p+3}$ can be defined as follows:
\[
s_{p,p+3}
\quad = \quad
\m{\begin{qcircuit}[scale = 0.4]
\grid{4}{0,1,2,3}
\leftlabel{\footnotesize $p$}{0,3}
\leftlabel{\footnotesize $p+1$}{0,2}
\leftlabel{\footnotesize $p+2$}{0,1}
\leftlabel{\footnotesize $p+3$}{0,0}
\dotswapgate{1}{0}{1}
\dotswapgate{1}{2}{3}
\dotswapgate{2}{1}{2}
\dotswapgate{3}{0}{1}
\dotswapgate{3}{2}{3}
\end{qcircuit}}
\]
Every fermionic permutation is a composition of fermionic
transpositions $s_{p,p+1}$. To decrease the run time, we should
minimize the depth of these permutation circuits.

The terms of the Hamiltonian can be classified according to the number
of distinct orbitals that they involve. That number may be
$1, 2, 3, 4$. We refer to these terms as singleton, pair, triple, and
quad terms, respectively. We can implement all the singleton terms in
parallel. For pair terms, it is natural to use Kirkman's circle method
for round-robin tournaments {\cite{Kirkman1847}}. This method ensures
that we implement these terms in a maximally time-efficient way and
that we minimize the time spent on swapping the orbitals to arrange
each new pairing.

Kirkman's circle method schedules a round-robin tournament for an even
number of competitors by fixing one of those competitors and cycling
the remaining competitors. We may imagine the circle method as a ski
lift with one skier, who is not on the ski lift, at the bottom of the
hill and the other skiers seated on the ski lift as it cycles.  At
each stage of the operation, we pair skiers seated across from each
other, and we pair the skier seated at the bottom with the stationary
skier:
\[
\begin{tikzpicture}[scale=0.4]
  \def\b{(0.7*pi)}
  \def\a{((7*\b - 2*pi) / 4)}
  \def\c{(5/4*\b)}
  \fill[fill=blue!20] (-0.5,{-\a-1+0.4}) rectangle (0.5,{-\a-3-0.4});
  \fill[fill=blue!20] (-1.25,{\c-0.5}) rectangle (1.25,{\c+0.5});
  \fill[fill=blue!20] (-1.25,{\c-\b-0.5}) rectangle (1.25,{\c-\b+0.5});
  \fill[fill=blue!20] (-1.25,{\c-2*\b-0.5}) rectangle (1.25,{\c-2*\b+0.5});
  \draw (1,0) -- (1,{\a}) .. controls (1,{\a+0.555}) and (0.555,{\a+1}) .. (0,{\a+1})
  .. controls (-0.555,{\a+1}) and (-1,{\a+0.555}) .. (-1,{\a})
  -- (-1,0) -- (-1,{-\a}) .. controls (-1,{-\a-0.555}) and (-0.555,{-\a-1}) .. (0,{-\a-1})
  .. controls (0.555,{-\a-1}) and (1,{-\a-0.555}) .. (1,{-\a}) -- cycle;
  \fill (0,{-\a-1}) circle (.15) node[below] {$6$};
  \fill (1,{\c-2*\b}) circle (.15) node[right] {$4$};
  \fill (-1,{\c-2*\b}) circle (.15) node[left] {$5$};
  \fill (1,{\c-\b}) circle (.15) node[right] {$2$};
  \fill (-1,{\c-\b}) circle (.15) node[left] {$3$};
  \fill (1,{\c}) node[right] {$0$};
  \fill (-1,{\c}) node[left] {$1$};
  \fill ({cos(180/pi*(\c-\a)},{\a+sin(180/pi*(\c-\a)}) circle (.15);
  \fill ({-cos(180/pi*(\c-\a)},{\a+sin(180/pi*(\c-\a)}) circle (.15);
  \fill (0,{-\a-3}) circle (.15) node[right=1mm] {$7$};
  \draw[-{Latex[length=1.5mm,width=2mm]}] (1,{\c-2*\b}) -- (1,{\c-1.5*\b+0.22});
  \draw[-{Latex[length=1.5mm,width=2mm]}] (-1,{\c-1*\b}) -- (-1,{\c-1.5*\b-0.22});
\end{tikzpicture}
\]
Thus, if we have $2n$ skiers, the ski lift goes through $2n-1$ stages
of operation, and we see $n$ pairs at each stage. If we have $2n-1$
skiers, then we simply omit the stationary skier, and whoever is at
the bottom of the ski lift will not be paired in that round.

Here is an example showing the 7 stages of the circle method for $8$ orbitals:
\[
\scalebox{0.8}{
\begin{qcircuit}[scale=0.4]
    \grid{34.0}{0,1,2,3,4,5,6,7}
    \wirelabel{\footnotesize 0}{0.779069767441861,7};
    \wirelabel{\footnotesize 1}{0.779069767441861,6};
    \wirelabel{\footnotesize 2}{0.779069767441861,5};
    \wirelabel{\footnotesize 3}{0.779069767441861,4};
    \wirelabel{\footnotesize 4}{0.779069767441861,3};
    \wirelabel{\footnotesize 5}{0.779069767441861,2};
    \wirelabel{\footnotesize 6}{0.779069767441861,1};
    \wirelabel{\footnotesize 7}{0.779069767441861,0};
    \biggate{$P$}{2.290697674418604,6}{2.290697674418604,7};
    \biggate{$P$}{2.290697674418604,4}{2.290697674418604,5};
    \biggate{$P$}{2.290697674418604,2}{2.290697674418604,3};
    \biggate{$P$}{2.290697674418604,0}{2.290697674418604,1};
    \dotswapgate{3.550387596899225}{6}{5};
    \dotswapgate{3.550387596899225}{4}{3};
    \dotswapgate{3.550387596899225}{2}{1};
    \dotswapgate{4.55813953488372}{7}{6};
    \dotswapgate{4.55813953488372}{5}{4};
    \dotswapgate{4.55813953488372}{3}{2};
    \wirelabel{\footnotesize 2}{5.817829457364342,7};
    \wirelabel{\footnotesize 0}{5.817829457364342,6};
    \wirelabel{\footnotesize 4}{5.817829457364342,5};
    \wirelabel{\footnotesize 1}{5.817829457364342,4};
    \wirelabel{\footnotesize 6}{5.817829457364342,3};
    \wirelabel{\footnotesize 3}{5.817829457364342,2};
    \wirelabel{\footnotesize 5}{5.817829457364342,1};
    \wirelabel{\footnotesize 7}{5.817829457364342,0};
    \biggate{$P$}{7.329457364341085,6}{7.329457364341085,7};
    \biggate{$P$}{7.329457364341085,4}{7.329457364341085,5};
    \biggate{$P$}{7.329457364341085,2}{7.329457364341085,3};
    \biggate{$P$}{7.329457364341085,0}{7.329457364341085,1};
    \dotswapgate{8.589147286821705}{6}{5};
    \dotswapgate{8.589147286821705}{4}{3};
    \dotswapgate{8.589147286821705}{2}{1};
    \dotswapgate{9.596899224806203}{7}{6};
    \dotswapgate{9.596899224806203}{5}{4};
    \dotswapgate{9.596899224806203}{3}{2};
    \wirelabel{\footnotesize 4}{10.856589147286822,7};
    \wirelabel{\footnotesize 2}{10.856589147286822,6};
    \wirelabel{\footnotesize 6}{10.856589147286822,5};
    \wirelabel{\footnotesize 0}{10.856589147286822,4};
    \wirelabel{\footnotesize 5}{10.856589147286822,3};
    \wirelabel{\footnotesize 1}{10.856589147286822,2};
    \wirelabel{\footnotesize 3}{10.856589147286822,1};
    \wirelabel{\footnotesize 7}{10.856589147286822,0};
    \biggate{$P$}{12.368217054263566,6}{12.368217054263566,7};
    \biggate{$P$}{12.368217054263566,4}{12.368217054263566,5};
    \biggate{$P$}{12.368217054263566,2}{12.368217054263566,3};
    \biggate{$P$}{12.368217054263566,0}{12.368217054263566,1};
    \dotswapgate{13.627906976744185}{6}{5};
    \dotswapgate{13.627906976744185}{4}{3};
    \dotswapgate{13.627906976744185}{2}{1};
    \dotswapgate{14.63565891472868}{7}{6};
    \dotswapgate{14.63565891472868}{5}{4};
    \dotswapgate{14.63565891472868}{3}{2};
    \wirelabel{\footnotesize 6}{15.895348837209305,7};
    \wirelabel{\footnotesize 4}{15.895348837209305,6};
    \wirelabel{\footnotesize 5}{15.895348837209305,5};
    \wirelabel{\footnotesize 2}{15.895348837209305,4};
    \wirelabel{\footnotesize 3}{15.895348837209305,3};
    \wirelabel{\footnotesize 0}{15.895348837209305,2};
    \wirelabel{\footnotesize 1}{15.895348837209305,1};
    \wirelabel{\footnotesize 7}{15.895348837209305,0};
    \biggate{$P$}{17.406976744186046,6}{17.406976744186046,7};
    \biggate{$P$}{17.406976744186046,4}{17.406976744186046,5};
    \biggate{$P$}{17.406976744186046,2}{17.406976744186046,3};
    \biggate{$P$}{17.406976744186046,0}{17.406976744186046,1};
    \dotswapgate{18.666666666666664}{6}{5};
    \dotswapgate{18.666666666666664}{4}{3};
    \dotswapgate{18.666666666666664}{2}{1};
    \dotswapgate{19.674418604651166}{7}{6};
    \dotswapgate{19.674418604651166}{5}{4};
    \dotswapgate{19.674418604651166}{3}{2};
    \wirelabel{\footnotesize 5}{20.934108527131784,7};
    \wirelabel{\footnotesize 6}{20.934108527131784,6};
    \wirelabel{\footnotesize 3}{20.934108527131784,5};
    \wirelabel{\footnotesize 4}{20.934108527131784,4};
    \wirelabel{\footnotesize 1}{20.934108527131784,3};
    \wirelabel{\footnotesize 2}{20.934108527131784,2};
    \wirelabel{\footnotesize 0}{20.934108527131784,1};
    \wirelabel{\footnotesize 7}{20.934108527131784,0};
    \biggate{$P$}{22.445736434108525,6}{22.445736434108525,7};
    \biggate{$P$}{22.445736434108525,4}{22.445736434108525,5};
    \biggate{$P$}{22.445736434108525,2}{22.445736434108525,3};
    \biggate{$P$}{22.445736434108525,0}{22.445736434108525,1};
    \dotswapgate{23.70542635658915}{6}{5};
    \dotswapgate{23.70542635658915}{4}{3};
    \dotswapgate{23.70542635658915}{2}{1};
    \dotswapgate{24.713178294573645}{7}{6};
    \dotswapgate{24.713178294573645}{5}{4};
    \dotswapgate{24.713178294573645}{3}{2};
    \wirelabel{\footnotesize 3}{25.972868217054263,7};
    \wirelabel{\footnotesize 5}{25.972868217054263,6};
    \wirelabel{\footnotesize 1}{25.972868217054263,5};
    \wirelabel{\footnotesize 6}{25.972868217054263,4};
    \wirelabel{\footnotesize 0}{25.972868217054263,3};
    \wirelabel{\footnotesize 4}{25.972868217054263,2};
    \wirelabel{\footnotesize 2}{25.972868217054263,1};
    \wirelabel{\footnotesize 7}{25.972868217054263,0};
    \biggate{$P$}{27.48449612403101,6}{27.48449612403101,7};
    \biggate{$P$}{27.48449612403101,4}{27.48449612403101,5};
    \biggate{$P$}{27.48449612403101,2}{27.48449612403101,3};
    \biggate{$P$}{27.48449612403101,0}{27.48449612403101,1};
    \dotswapgate{28.74418604651163}{6}{5};
    \dotswapgate{28.74418604651163}{4}{3};
    \dotswapgate{28.74418604651163}{2}{1};
    \dotswapgate{29.751937984496124}{7}{6};
    \dotswapgate{29.751937984496124}{5}{4};
    \dotswapgate{29.751937984496124}{3}{2};
    \wirelabel{\footnotesize 1}{31.01162790697675,7};
    \wirelabel{\footnotesize 3}{31.01162790697675,6};
    \wirelabel{\footnotesize 0}{31.01162790697675,5};
    \wirelabel{\footnotesize 5}{31.01162790697675,4};
    \wirelabel{\footnotesize 2}{31.01162790697675,3};
    \wirelabel{\footnotesize 6}{31.01162790697675,2};
    \wirelabel{\footnotesize 4}{31.01162790697675,1};
    \wirelabel{\footnotesize 7}{31.01162790697675,0};
    \biggate{$P$}{32.52325581395349,6}{32.52325581395349,7};
    \biggate{$P$}{32.52325581395349,4}{32.52325581395349,5};
    \biggate{$P$}{32.52325581395349,2}{32.52325581395349,3};
    \biggate{$P$}{32.52325581395349,0}{32.52325581395349,1};
\end{qcircuit}}
\]

We solve the scheduling problem for triple terms by a shrewd
application of M\"{o}bius transformations over finite fields. We find
the smallest prime $p \geq m - 1$. Recall that $\FF_p$ is the field of
integers modulo $p$, and that a M\"{o}bius transformation is a
permutation of $\FF_p \cup \{\infty\}$ of the form
$z \mapsto (a z + b)/ (c z + d)$ for parameters
$a, b, c, d \in \FF_p$. Each M\"{o}bius transformation of order three
partitions $\FF_p \cup \{\infty\}$ into orbits, at most two of which
are singletons. Furthermore, every subset of cardinality three is an
orbit of exactly two such M\"{o}bius transformations, which are each
other's inverses. Thus, if $m = p+1$, then we obtain a sequence of
partitions of our orbital qubits into triples such that every triple
is in exactly one partition. We obtain maximal parallelization for the
implementation of these Hamiltonian terms.  If $m \neq p + 1$, then we
simply ignore partition blocks that contain elements greater than
$m-1$.

We solve the scheduling problem for quad terms by combining the ski
lift method with M\"{o}bius transformations. The intuition is that
each seat of the ski lift will now hold two skiers instead of just
one. A single cycle thus implements all possible quadruples of orbital
qubits that may be obtained from a single partition of the orbital
qubits into pairs. We use M\"{o}bius transformations to find a small
set of such partitions that implements every quadruple, applying an
idea of Nazarov and Speyer {\cite{Nazarov-Speyer}}. Specifically, we
use M\"{o}bius transformations of order two to obtain partitions of
$\FF_p \cup \{\infty\}$ into pairs. Each quadruple is implemented by
partitions arising from three such M\"{o}bius transformations. By
considering the composition of these three M\"{o}bius transformations,
we conclude that it is sufficient to use only those M\"{o}bius
transformations whose determinant $ad-bc$ is a quadratic residue
modulo $p$. As in the case of triples, we ignore partition blocks that contain
elements greater than $m-1$ in the common case that $m \neq p + 1$.

% ----------------------------------------------------------------------
\section{Putting everything together}

\begin{figure}[p]
  \def\v{\vspace{-2em}}
  (a)\v
  \[ \includegraphics[width=0.7\textwidth]{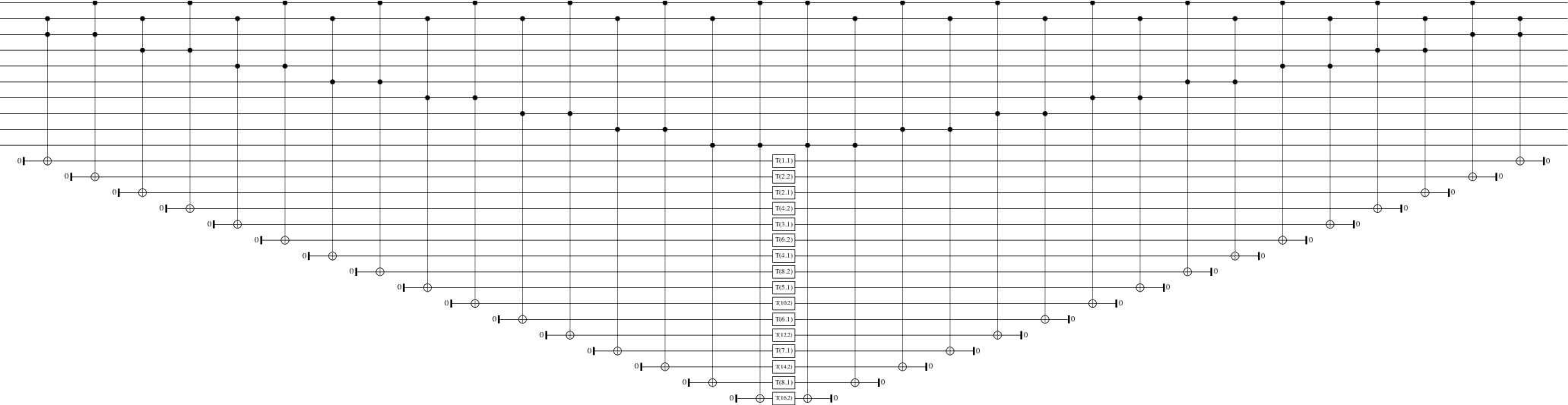}
  \]
  (b)\v
  \[ \includegraphics[width=0.9\textwidth]{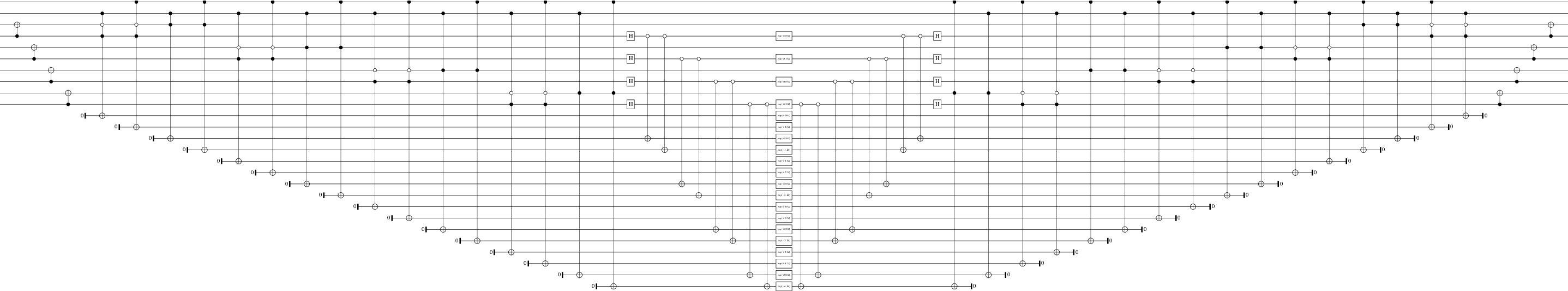}
  \]
  (c)\v
  \[ \includegraphics[width=0.9\textwidth]{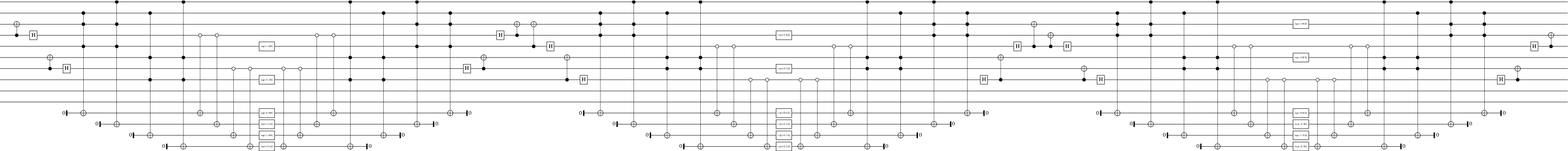}
  \]
  (d)\v
  \[ \includegraphics[width=0.9\textwidth]{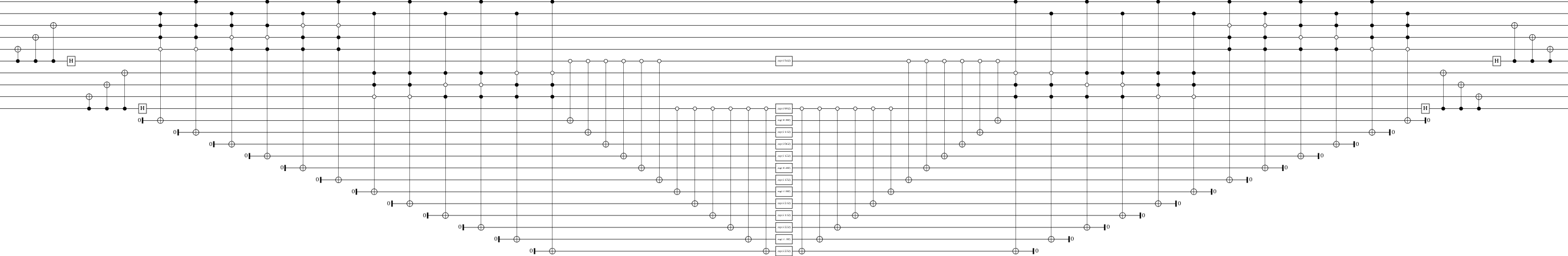}
  \]
  \caption{A typical singleton, pair, triple, and quad stage with
    $b=2$ precision qubits and $m=8$ orbital qubits. The topmost $2$
    inputs are the precision qubits and the bottommost $8$ inputs are
    the orbital qubits. In (a), (b), and (d), note that all
    single-qubit rotations are done in parallel, using the method of
    Section~\ref{ssec:parallel-controlled-rotations} to decompose
    controlled rotations. The circuit (c) requires 3 rounds of
    rotations, because for each triple of orbitals $(p,q,r)$, there are
    Hamiltonian terms of the forms
    $a\da_p a\da_q a_q a_r + a\da_r a\da_q a_q a_p$,
    $a\da_q a\da_p a_p a_r + a\da_r a\da_p a_p a_q$, and
    $a\da_p a\da_r a_r a_q + a\da_q a\da_r a_r a_p$. Since these
    do not commute with each other, they cannot be performed in
    parallel. By contrast, the circuit (d) requires only one round of
    rotations, because Hamiltonian quad terms on $(p,q,r,s)$ are of
    the forms
    $a\da_p a\da_q a_r a_s + a\da_s a\da_r a_q a_p$,
    $a\da_p a\da_r a_q a_s + a\da_s a\da_q a_r a_p$, and
    $a\da_p a\da_s a_r a_q + a\da_q a\da_r a_s a_p$.
    These terms do commute with each other, and
    therefore can be performed in parallel.
    In (a)--(d), the multiply-controlled not gates that are used to
    prepare and uncompute the various ancillas can all be realized in
    constant time in the lattice surgery framework, i.e., in time that
    is independent of both $b$ and $m$.
  }
  \label{fig-stages}
\end{figure}

\begin{figure}
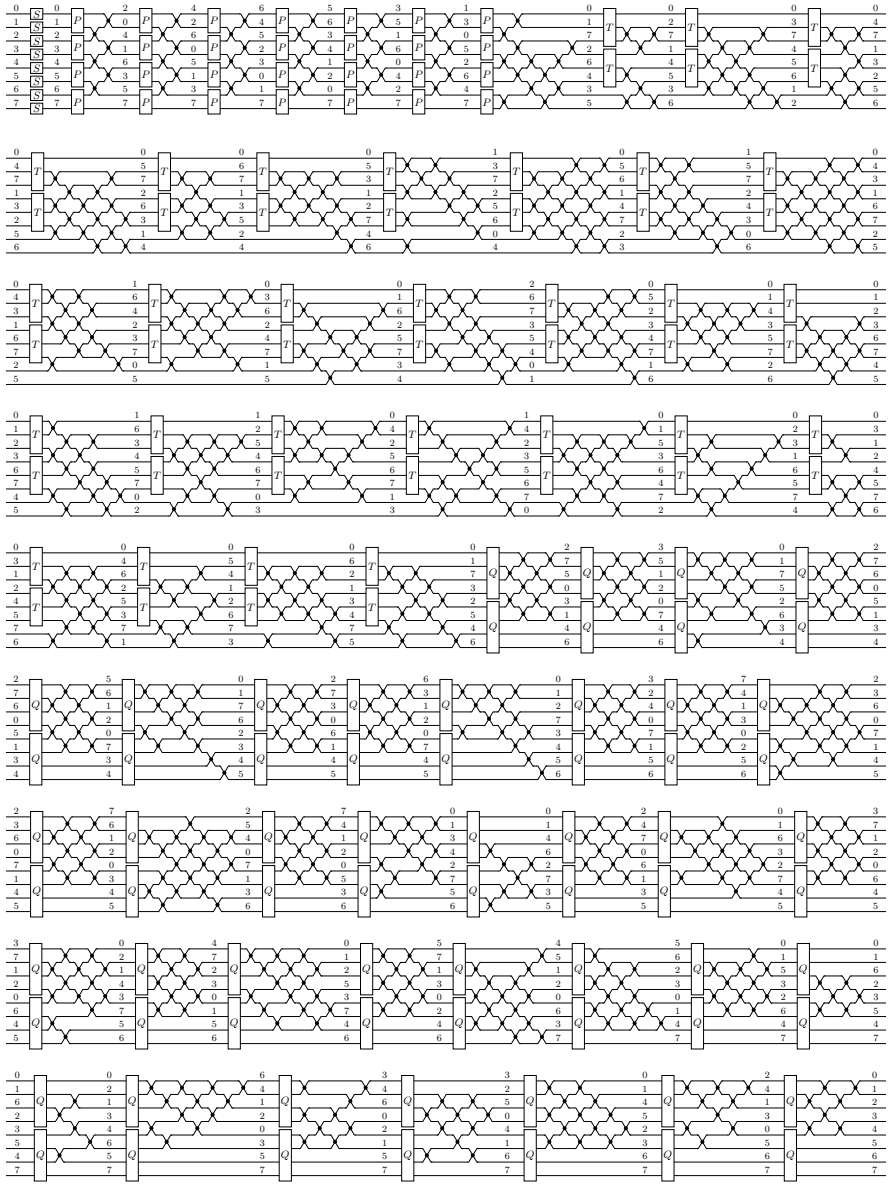

\[
  \scalebox{0.45}{
  \begin{qcircuit}[scale=0.4]
    \grid{65.0}{0,1,2,3,4,5,6,7}
    \wirelabel{\footnotesize 0}{0.7558139534883721,7};
    \wirelabel{\footnotesize 1}{0.7558139534883721,6};
    \wirelabel{\footnotesize 2}{0.7558139534883721,5};
    \wirelabel{\footnotesize 3}{0.7558139534883721,4};
    \wirelabel{\footnotesize 4}{0.7558139534883721,3};
    \wirelabel{\footnotesize 5}{0.7558139534883721,2};
    \wirelabel{\footnotesize 6}{0.7558139534883721,1};
    \wirelabel{\footnotesize 7}{0.7558139534883721,0};
    \gate{$S$}{2.2674418604651163,7};
    \gate{$S$}{2.2674418604651163,6};
    \gate{$S$}{2.2674418604651163,5};
    \gate{$S$}{2.2674418604651163,4};
    \gate{$S$}{2.2674418604651163,3};
    \gate{$S$}{2.2674418604651163,2};
    \gate{$S$}{2.2674418604651163,1};
    \gate{$S$}{2.2674418604651163,0};
    \wirelabel{\footnotesize 0}{3.779069767441861,7};
    \wirelabel{\footnotesize 1}{3.779069767441861,6};
    \wirelabel{\footnotesize 2}{3.779069767441861,5};
    \wirelabel{\footnotesize 3}{3.779069767441861,4};
    \wirelabel{\footnotesize 4}{3.779069767441861,3};
    \wirelabel{\footnotesize 5}{3.779069767441861,2};
    \wirelabel{\footnotesize 6}{3.779069767441861,1};
    \wirelabel{\footnotesize 7}{3.779069767441861,0};
    \biggate{$P$}{5.290697674418604,6}{5.290697674418604,7};
    \biggate{$P$}{5.290697674418604,4}{5.290697674418604,5};
    \biggate{$P$}{5.290697674418604,2}{5.290697674418604,3};
    \biggate{$P$}{5.290697674418604,0}{5.290697674418604,1};
    \dotswapgate{6.550387596899225}{6}{5};
    \dotswapgate{6.550387596899225}{4}{3};
    \dotswapgate{6.550387596899225}{2}{1};
    \dotswapgate{7.55813953488372}{7}{6};
    \dotswapgate{7.55813953488372}{5}{4};
    \dotswapgate{7.55813953488372}{3}{2};
    \wirelabel{\footnotesize 2}{8.817829457364342,7};
    \wirelabel{\footnotesize 0}{8.817829457364342,6};
    \wirelabel{\footnotesize 4}{8.817829457364342,5};
    \wirelabel{\footnotesize 1}{8.817829457364342,4};
    \wirelabel{\footnotesize 6}{8.817829457364342,3};
    \wirelabel{\footnotesize 3}{8.817829457364342,2};
    \wirelabel{\footnotesize 5}{8.817829457364342,1};
    \wirelabel{\footnotesize 7}{8.817829457364342,0};
    \biggate{$P$}{10.329457364341085,6}{10.329457364341085,7};
    \biggate{$P$}{10.329457364341085,4}{10.329457364341085,5};
    \biggate{$P$}{10.329457364341085,2}{10.329457364341085,3};
    \biggate{$P$}{10.329457364341085,0}{10.329457364341085,1};
    \dotswapgate{11.589147286821705}{6}{5};
    \dotswapgate{11.589147286821705}{4}{3};
    \dotswapgate{11.589147286821705}{2}{1};
    \dotswapgate{12.596899224806203}{7}{6};
    \dotswapgate{12.596899224806203}{5}{4};
    \dotswapgate{12.596899224806203}{3}{2};
    \wirelabel{\footnotesize 4}{13.856589147286822,7};
    \wirelabel{\footnotesize 2}{13.856589147286822,6};
    \wirelabel{\footnotesize 6}{13.856589147286822,5};
    \wirelabel{\footnotesize 0}{13.856589147286822,4};
    \wirelabel{\footnotesize 5}{13.856589147286822,3};
    \wirelabel{\footnotesize 1}{13.856589147286822,2};
    \wirelabel{\footnotesize 3}{13.856589147286822,1};
    \wirelabel{\footnotesize 7}{13.856589147286822,0};
    \biggate{$P$}{15.368217054263566,6}{15.368217054263566,7};
    \biggate{$P$}{15.368217054263566,4}{15.368217054263566,5};
    \biggate{$P$}{15.368217054263566,2}{15.368217054263566,3};
    \biggate{$P$}{15.368217054263566,0}{15.368217054263566,1};
    \dotswapgate{16.627906976744185}{6}{5};
    \dotswapgate{16.627906976744185}{4}{3};
    \dotswapgate{16.627906976744185}{2}{1};
    \dotswapgate{17.63565891472868}{7}{6};
    \dotswapgate{17.63565891472868}{5}{4};
    \dotswapgate{17.63565891472868}{3}{2};
    \wirelabel{\footnotesize 6}{18.895348837209305,7};
    \wirelabel{\footnotesize 4}{18.895348837209305,6};
    \wirelabel{\footnotesize 5}{18.895348837209305,5};
    \wirelabel{\footnotesize 2}{18.895348837209305,4};
    \wirelabel{\footnotesize 3}{18.895348837209305,3};
    \wirelabel{\footnotesize 0}{18.895348837209305,2};
    \wirelabel{\footnotesize 1}{18.895348837209305,1};
    \wirelabel{\footnotesize 7}{18.895348837209305,0};
    \biggate{$P$}{20.406976744186046,6}{20.406976744186046,7};
    \biggate{$P$}{20.406976744186046,4}{20.406976744186046,5};
    \biggate{$P$}{20.406976744186046,2}{20.406976744186046,3};
    \biggate{$P$}{20.406976744186046,0}{20.406976744186046,1};
    \dotswapgate{21.666666666666664}{6}{5};
    \dotswapgate{21.666666666666664}{4}{3};
    \dotswapgate{21.666666666666664}{2}{1};
    \dotswapgate{22.674418604651166}{7}{6};
    \dotswapgate{22.674418604651166}{5}{4};
    \dotswapgate{22.674418604651166}{3}{2};
    \wirelabel{\footnotesize 5}{23.934108527131784,7};
    \wirelabel{\footnotesize 6}{23.934108527131784,6};
    \wirelabel{\footnotesize 3}{23.934108527131784,5};
    \wirelabel{\footnotesize 4}{23.934108527131784,4};
    \wirelabel{\footnotesize 1}{23.934108527131784,3};
    \wirelabel{\footnotesize 2}{23.934108527131784,2};
    \wirelabel{\footnotesize 0}{23.934108527131784,1};
    \wirelabel{\footnotesize 7}{23.934108527131784,0};
    \biggate{$P$}{25.445736434108525,6}{25.445736434108525,7};
    \biggate{$P$}{25.445736434108525,4}{25.445736434108525,5};
    \biggate{$P$}{25.445736434108525,2}{25.445736434108525,3};
    \biggate{$P$}{25.445736434108525,0}{25.445736434108525,1};
    \dotswapgate{26.70542635658915}{6}{5};
    \dotswapgate{26.70542635658915}{4}{3};
    \dotswapgate{26.70542635658915}{2}{1};
    \dotswapgate{27.713178294573645}{7}{6};
    \dotswapgate{27.713178294573645}{5}{4};
    \dotswapgate{27.713178294573645}{3}{2};
    \wirelabel{\footnotesize 3}{28.972868217054263,7};
    \wirelabel{\footnotesize 5}{28.972868217054263,6};
    \wirelabel{\footnotesize 1}{28.972868217054263,5};
    \wirelabel{\footnotesize 6}{28.972868217054263,4};
    \wirelabel{\footnotesize 0}{28.972868217054263,3};
    \wirelabel{\footnotesize 4}{28.972868217054263,2};
    \wirelabel{\footnotesize 2}{28.972868217054263,1};
    \wirelabel{\footnotesize 7}{28.972868217054263,0};
    \biggate{$P$}{30.48449612403101,6}{30.48449612403101,7};
    \biggate{$P$}{30.48449612403101,4}{30.48449612403101,5};
    \biggate{$P$}{30.48449612403101,2}{30.48449612403101,3};
    \biggate{$P$}{30.48449612403101,0}{30.48449612403101,1};
    \dotswapgate{31.74418604651163}{6}{5};
    \dotswapgate{31.74418604651163}{4}{3};
    \dotswapgate{31.74418604651163}{2}{1};
    \dotswapgate{32.751937984496124}{7}{6};
    \dotswapgate{32.751937984496124}{5}{4};
    \dotswapgate{32.751937984496124}{3}{2};
    \wirelabel{\footnotesize 1}{34.01162790697675,7};
    \wirelabel{\footnotesize 3}{34.01162790697675,6};
    \wirelabel{\footnotesize 0}{34.01162790697675,5};
    \wirelabel{\footnotesize 5}{34.01162790697675,4};
    \wirelabel{\footnotesize 2}{34.01162790697675,3};
    \wirelabel{\footnotesize 6}{34.01162790697675,2};
    \wirelabel{\footnotesize 4}{34.01162790697675,1};
    \wirelabel{\footnotesize 7}{34.01162790697675,0};
    \biggate{$P$}{35.52325581395349,6}{35.52325581395349,7};
    \biggate{$P$}{35.52325581395349,4}{35.52325581395349,5};
    \biggate{$P$}{35.52325581395349,2}{35.52325581395349,3};
    \biggate{$P$}{35.52325581395349,0}{35.52325581395349,1};
    \dotswapgate{36.78294573643411}{6}{5};
    \dotswapgate{36.78294573643411}{4}{3};
    \dotswapgate{36.78294573643411}{1}{0};
    \dotswapgate{37.79069767441861}{7}{6};
    \dotswapgate{37.79069767441861}{5}{4};
    \dotswapgate{37.79069767441861}{3}{2};
    \dotswapgate{38.798449612403104}{4}{3};
    \dotswapgate{38.798449612403104}{2}{1};
    \dotswapgate{39.8062015503876}{3}{2};
    \dotswapgate{39.8062015503876}{1}{0};
    \dotswapgate{40.81395348837209}{4}{3};
    \dotswapgate{40.81395348837209}{2}{1};
    \dotswapgate{41.821705426356594}{5}{4};
    \wirelabel{\footnotesize 0}{43.08139534883721,7};
    \wirelabel{\footnotesize 1}{43.08139534883721,6};
    \wirelabel{\footnotesize 7}{43.08139534883721,5};
    \wirelabel{\footnotesize 2}{43.08139534883721,4};
    \wirelabel{\footnotesize 6}{43.08139534883721,3};
    \wirelabel{\footnotesize 4}{43.08139534883721,2};
    \wirelabel{\footnotesize 3}{43.08139534883721,1};
    \wirelabel{\footnotesize 5}{43.08139534883721,0};
    \biggate{$T$}{44.593023255813954,5}{44.593023255813954,7};
    \biggate{$T$}{44.593023255813954,2}{44.593023255813954,4};
    \dotswapgate{45.85271317829458}{6}{5};
    \dotswapgate{45.85271317829458}{3}{2};
    \dotswapgate{45.85271317829458}{1}{0};
    \dotswapgate{46.86046511627907}{5}{4};
    \dotswapgate{46.86046511627907}{2}{1};
    \dotswapgate{47.86821705426357}{6}{5};
    \dotswapgate{47.86821705426357}{1}{0};
    \wirelabel{\footnotesize 0}{49.127906976744185,7};
    \wirelabel{\footnotesize 2}{49.127906976744185,6};
    \wirelabel{\footnotesize 7}{49.127906976744185,5};
    \wirelabel{\footnotesize 1}{49.127906976744185,4};
    \wirelabel{\footnotesize 4}{49.127906976744185,3};
    \wirelabel{\footnotesize 5}{49.127906976744185,2};
    \wirelabel{\footnotesize 3}{49.127906976744185,1};
    \wirelabel{\footnotesize 6}{49.127906976744185,0};
    \biggate{$T$}{50.639534883720934,5}{50.639534883720934,7};
    \biggate{$T$}{50.639534883720934,2}{50.639534883720934,4};
    \dotswapgate{51.89922480620155}{6}{5};
    \dotswapgate{51.89922480620155}{4}{3};
    \dotswapgate{51.89922480620155}{2}{1};
    \dotswapgate{52.906976744186046}{5}{4};
    \dotswapgate{52.906976744186046}{3}{2};
    \dotswapgate{53.91472868217055}{4}{3};
    \dotswapgate{53.91472868217055}{2}{1};
    \dotswapgate{54.92248062015504}{5}{4};
    \dotswapgate{54.92248062015504}{3}{2};
    \dotswapgate{54.92248062015504}{1}{0};
    \dotswapgate{55.93023255813954}{6}{5};
    \dotswapgate{55.93023255813954}{2}{1};
    \dotswapgate{56.93798449612403}{1}{0};
    \wirelabel{\footnotesize 0}{58.197674418604656,7};
    \wirelabel{\footnotesize 3}{58.197674418604656,6};
    \wirelabel{\footnotesize 7}{58.197674418604656,5};
    \wirelabel{\footnotesize 4}{58.197674418604656,4};
    \wirelabel{\footnotesize 5}{58.197674418604656,3};
    \wirelabel{\footnotesize 6}{58.197674418604656,2};
    \wirelabel{\footnotesize 1}{58.197674418604656,1};
    \wirelabel{\footnotesize 2}{58.197674418604656,0};
    \biggate{$T$}{59.7093023255814,5}{59.7093023255814,7};
    \biggate{$T$}{59.7093023255814,2}{59.7093023255814,4};
    \dotswapgate{60.968992248062015}{6}{5};
    \dotswapgate{60.968992248062015}{2}{1};
    \dotswapgate{61.97674418604652}{5}{4};
    \dotswapgate{61.97674418604652}{3}{2};
    \dotswapgate{61.97674418604652}{1}{0};
    \dotswapgate{62.98449612403101}{6}{5};
    \dotswapgate{62.98449612403101}{4}{3};
    \dotswapgate{62.98449612403101}{2}{1};
    \wirelabel{\footnotesize 0}{64.24418604651163,7};
    \wirelabel{\footnotesize 4}{64.24418604651163,6};
    \wirelabel{\footnotesize 7}{64.24418604651163,5};
    \wirelabel{\footnotesize 1}{64.24418604651163,4};
    \wirelabel{\footnotesize 3}{64.24418604651163,3};
    \wirelabel{\footnotesize 2}{64.24418604651163,2};
    \wirelabel{\footnotesize 5}{64.24418604651163,1};
    \wirelabel{\footnotesize 6}{64.24418604651163,0};
  \end{qcircuit}
}
\]
\[
  \scalebox{0.45}{
  \begin{qcircuit}[scale=0.4]
    \grid{65.0}{0,1,2,3,4,5,6,7}
    \wirelabel{\footnotesize 0}{0.7800000000000011,7};
    \wirelabel{\footnotesize 4}{0.7800000000000011,6};
    \wirelabel{\footnotesize 7}{0.7800000000000011,5};
    \wirelabel{\footnotesize 1}{0.7800000000000011,4};
    \wirelabel{\footnotesize 3}{0.7800000000000011,3};
    \wirelabel{\footnotesize 2}{0.7800000000000011,2};
    \wirelabel{\footnotesize 5}{0.7800000000000011,1};
    \wirelabel{\footnotesize 6}{0.7800000000000011,0};
    \biggate{$T$}{2.3400000000000034,5}{2.3400000000000034,7};
    \biggate{$T$}{2.3400000000000034,2}{2.3400000000000034,4};
    \dotswapgate{3.6400000000000006}{6}{5};
    \dotswapgate{3.6400000000000006}{4}{3};
    \dotswapgate{3.6400000000000006}{2}{1};
    \dotswapgate{4.680000000000007}{5}{4};
    \dotswapgate{4.680000000000007}{3}{2};
    \dotswapgate{5.719999999999999}{4}{3};
    \dotswapgate{5.719999999999999}{2}{1};
    \dotswapgate{6.760000000000005}{5}{4};
    \dotswapgate{6.760000000000005}{3}{2};
    \dotswapgate{6.760000000000005}{1}{0};
    \dotswapgate{7.799999999999997}{6}{5};
    \dotswapgate{7.799999999999997}{4}{3};
    \dotswapgate{7.799999999999997}{2}{1};
    \dotswapgate{8.840000000000003}{3}{2};
    \dotswapgate{8.840000000000003}{1}{0};
    \wirelabel{\footnotesize 0}{10.14,7};
    \wirelabel{\footnotesize 5}{10.14,6};
    \wirelabel{\footnotesize 7}{10.14,5};
    \wirelabel{\footnotesize 2}{10.14,4};
    \wirelabel{\footnotesize 6}{10.14,3};
    \wirelabel{\footnotesize 3}{10.14,2};
    \wirelabel{\footnotesize 1}{10.14,1};
    \wirelabel{\footnotesize 4}{10.14,0};
    \biggate{$T$}{11.700000000000003,5}{11.700000000000003,7};
    \biggate{$T$}{11.700000000000003,2}{11.700000000000003,4};
    \dotswapgate{13.0}{6}{5};
    \dotswapgate{13.0}{4}{3};
    \dotswapgate{13.0}{2}{1};
    \dotswapgate{14.040000000000006}{5}{4};
    \dotswapgate{14.040000000000006}{3}{2};
    \dotswapgate{15.079999999999998}{6}{5};
    \dotswapgate{15.079999999999998}{4}{3};
    \dotswapgate{15.079999999999998}{2}{1};
    \dotswapgate{16.120000000000005}{3}{2};
    \wirelabel{\footnotesize 0}{17.42,7};
    \wirelabel{\footnotesize 6}{17.42,6};
    \wirelabel{\footnotesize 7}{17.42,5};
    \wirelabel{\footnotesize 1}{17.42,4};
    \wirelabel{\footnotesize 3}{17.42,3};
    \wirelabel{\footnotesize 5}{17.42,2};
    \wirelabel{\footnotesize 2}{17.42,1};
    \wirelabel{\footnotesize 4}{17.42,0};
    \biggate{$T$}{18.980000000000004,5}{18.980000000000004,7};
    \biggate{$T$}{18.980000000000004,2}{18.980000000000004,4};
    \dotswapgate{20.28}{6}{5};
    \dotswapgate{20.28}{4}{3};
    \dotswapgate{21.320000000000007}{5}{4};
    \dotswapgate{21.320000000000007}{3}{2};
    \dotswapgate{22.36}{6}{5};
    \dotswapgate{22.36}{4}{3};
    \dotswapgate{23.400000000000006}{5}{4};
    \dotswapgate{23.400000000000006}{3}{2};
    \dotswapgate{24.439999999999998}{6}{5};
    \dotswapgate{24.439999999999998}{4}{3};
    \dotswapgate{24.439999999999998}{2}{1};
    \dotswapgate{25.480000000000004}{3}{2};
    \dotswapgate{25.480000000000004}{1}{0};
    \wirelabel{\footnotesize 0}{26.78,7};
    \wirelabel{\footnotesize 5}{26.78,6};
    \wirelabel{\footnotesize 3}{26.78,5};
    \wirelabel{\footnotesize 1}{26.78,4};
    \wirelabel{\footnotesize 2}{26.78,3};
    \wirelabel{\footnotesize 7}{26.78,2};
    \wirelabel{\footnotesize 4}{26.78,1};
    \wirelabel{\footnotesize 6}{26.78,0};
    \biggate{$T$}{28.340000000000003,5}{28.340000000000003,7};
    \biggate{$T$}{28.340000000000003,2}{28.340000000000003,4};
    \dotswapgate{29.64}{7}{6};
    \dotswapgate{29.64}{5}{4};
    \dotswapgate{29.64}{3}{2};
    \dotswapgate{29.64}{1}{0};
    \dotswapgate{30.680000000000007}{6}{5};
    \dotswapgate{31.72}{7}{6};
    \dotswapgate{31.72}{5}{4};
    \dotswapgate{32.760000000000005}{6}{5};
    \dotswapgate{32.760000000000005}{4}{3};
    \dotswapgate{33.8}{5}{4};
    \dotswapgate{33.8}{3}{2};
    \dotswapgate{34.84}{4}{3};
    \dotswapgate{34.84}{2}{1};
    \wirelabel{\footnotesize 1}{36.14,7};
    \wirelabel{\footnotesize 3}{36.14,6};
    \wirelabel{\footnotesize 7}{36.14,5};
    \wirelabel{\footnotesize 2}{36.14,4};
    \wirelabel{\footnotesize 5}{36.14,3};
    \wirelabel{\footnotesize 6}{36.14,2};
    \wirelabel{\footnotesize 0}{36.14,1};
    \wirelabel{\footnotesize 4}{36.14,0};
    \biggate{$T$}{37.7,5}{37.7,7};
    \biggate{$T$}{37.7,2}{37.7,4};
    \dotswapgate{39.0}{6}{5};
    \dotswapgate{39.0}{4}{3};
    \dotswapgate{39.0}{2}{1};
    \dotswapgate{40.040000000000006}{5}{4};
    \dotswapgate{40.040000000000006}{3}{2};
    \dotswapgate{41.08}{6}{5};
    \dotswapgate{41.08}{4}{3};
    \dotswapgate{41.08}{2}{1};
    \dotswapgate{42.120000000000005}{7}{6};
    \dotswapgate{42.120000000000005}{5}{4};
    \dotswapgate{42.120000000000005}{3}{2};
    \dotswapgate{42.120000000000005}{1}{0};
    \dotswapgate{43.16}{6}{5};
    \dotswapgate{43.16}{4}{3};
    \dotswapgate{43.16}{2}{1};
    \dotswapgate{44.2}{7}{6};
    \dotswapgate{44.2}{5}{4};
    \dotswapgate{44.2}{3}{2};
    \dotswapgate{44.2}{1}{0};
    \wirelabel{\footnotesize 0}{45.5,7};
    \wirelabel{\footnotesize 5}{45.5,6};
    \wirelabel{\footnotesize 6}{45.5,5};
    \wirelabel{\footnotesize 1}{45.5,4};
    \wirelabel{\footnotesize 4}{45.5,3};
    \wirelabel{\footnotesize 7}{45.5,2};
    \wirelabel{\footnotesize 2}{45.5,1};
    \wirelabel{\footnotesize 3}{45.5,0};
    \biggate{$T$}{47.06,5}{47.06,7};
    \biggate{$T$}{47.06,2}{47.06,4};
    \dotswapgate{48.36}{7}{6};
    \dotswapgate{48.36}{5}{4};
    \dotswapgate{48.36}{3}{2};
    \dotswapgate{49.400000000000006}{6}{5};
    \dotswapgate{49.400000000000006}{4}{3};
    \dotswapgate{49.400000000000006}{2}{1};
    \dotswapgate{50.44}{7}{6};
    \dotswapgate{50.44}{5}{4};
    \dotswapgate{50.44}{3}{2};
    \dotswapgate{51.480000000000004}{4}{3};
    \dotswapgate{51.480000000000004}{2}{1};
    \dotswapgate{52.519999999999996}{3}{2};
    \dotswapgate{52.519999999999996}{1}{0};
    \dotswapgate{53.56}{2}{1};
    \wirelabel{\footnotesize 1}{54.86,7};
    \wirelabel{\footnotesize 5}{54.86,6};
    \wirelabel{\footnotesize 7}{54.86,5};
    \wirelabel{\footnotesize 2}{54.86,4};
    \wirelabel{\footnotesize 4}{54.86,3};
    \wirelabel{\footnotesize 3}{54.86,2};
    \wirelabel{\footnotesize 0}{54.86,1};
    \wirelabel{\footnotesize 6}{54.86,0};
    \biggate{$T$}{56.42,5}{56.42,7};
    \biggate{$T$}{56.42,2}{56.42,4};
    \dotswapgate{57.72}{6}{5};
    \dotswapgate{57.72}{4}{3};
    \dotswapgate{57.72}{2}{1};
    \dotswapgate{58.760000000000005}{5}{4};
    \dotswapgate{58.760000000000005}{3}{2};
    \dotswapgate{59.8}{6}{5};
    \dotswapgate{59.8}{4}{3};
    \dotswapgate{59.8}{2}{1};
    \dotswapgate{60.84}{7}{6};
    \dotswapgate{60.84}{5}{4};
    \dotswapgate{60.84}{3}{2};
    \dotswapgate{60.84}{1}{0};
    \dotswapgate{61.879999999999995}{6}{5};
    \dotswapgate{61.879999999999995}{4}{3};
    \dotswapgate{61.879999999999995}{2}{1};
    \dotswapgate{62.920000000000016}{7}{6};
    \dotswapgate{62.920000000000016}{5}{4};
    \dotswapgate{62.920000000000016}{3}{2};
    \dotswapgate{62.920000000000016}{1}{0};
    \wirelabel{\footnotesize 0}{64.22,7};
    \wirelabel{\footnotesize 4}{64.22,6};
    \wirelabel{\footnotesize 3}{64.22,5};
    \wirelabel{\footnotesize 1}{64.22,4};
    \wirelabel{\footnotesize 6}{64.22,3};
    \wirelabel{\footnotesize 7}{64.22,2};
    \wirelabel{\footnotesize 2}{64.22,1};
    \wirelabel{\footnotesize 5}{64.22,0};
  \end{qcircuit}
}
\]
\[
  \scalebox{0.45}{
  \begin{qcircuit}[scale=0.4]
    \grid{65.0}{0,1,2,3,4,5,6,7}
    \wirelabel{\footnotesize 0}{0.7330827067669219,7};
    \wirelabel{\footnotesize 4}{0.7330827067669219,6};
    \wirelabel{\footnotesize 3}{0.7330827067669219,5};
    \wirelabel{\footnotesize 1}{0.7330827067669219,4};
    \wirelabel{\footnotesize 6}{0.7330827067669219,3};
    \wirelabel{\footnotesize 7}{0.7330827067669219,2};
    \wirelabel{\footnotesize 2}{0.7330827067669219,1};
    \wirelabel{\footnotesize 5}{0.7330827067669219,0};
    \biggate{$T$}{2.1992481203007515,5}{2.1992481203007515,7};
    \biggate{$T$}{2.1992481203007515,2}{2.1992481203007515,4};
    \dotswapgate{3.421052631578945}{7}{6};
    \dotswapgate{3.421052631578945}{5}{4};
    \dotswapgate{3.421052631578945}{2}{1};
    \dotswapgate{4.398496240601503}{6}{5};
    \dotswapgate{4.398496240601503}{4}{3};
    \dotswapgate{5.375939849624061}{7}{6};
    \dotswapgate{5.375939849624061}{5}{4};
    \dotswapgate{5.375939849624061}{3}{2};
    \dotswapgate{6.3533834586466185}{6}{5};
    \dotswapgate{6.3533834586466185}{4}{3};
    \dotswapgate{7.3308270676691905}{3}{2};
    \dotswapgate{8.308270676691734}{2}{1};
    \wirelabel{\footnotesize 1}{9.530075187969942,7};
    \wirelabel{\footnotesize 6}{9.530075187969942,6};
    \wirelabel{\footnotesize 4}{9.530075187969942,5};
    \wirelabel{\footnotesize 2}{9.530075187969942,4};
    \wirelabel{\footnotesize 3}{9.530075187969942,3};
    \wirelabel{\footnotesize 7}{9.530075187969942,2};
    \wirelabel{\footnotesize 0}{9.530075187969942,1};
    \wirelabel{\footnotesize 5}{9.530075187969942,0};
    \biggate{$T$}{10.996240601503757,5}{10.996240601503757,7};
    \biggate{$T$}{10.996240601503757,2}{10.996240601503757,4};
    \dotswapgate{12.218045112781965}{7}{6};
    \dotswapgate{12.218045112781965}{5}{4};
    \dotswapgate{12.218045112781965}{2}{1};
    \dotswapgate{13.195488721804509}{6}{5};
    \dotswapgate{13.195488721804509}{4}{3};
    \dotswapgate{14.17293233082708}{5}{4};
    \dotswapgate{14.17293233082708}{3}{2};
    \dotswapgate{15.150375939849624}{6}{5};
    \dotswapgate{15.150375939849624}{4}{3};
    \dotswapgate{16.127819548872196}{7}{6};
    \dotswapgate{16.127819548872196}{5}{4};
    \dotswapgate{16.127819548872196}{3}{2};
    \dotswapgate{17.10526315789474}{6}{5};
    \dotswapgate{17.10526315789474}{2}{1};
    \dotswapgate{18.082706766917283}{7}{6};
    \wirelabel{\footnotesize 0}{19.30451127819549,7};
    \wirelabel{\footnotesize 3}{19.30451127819549,6};
    \wirelabel{\footnotesize 6}{19.30451127819549,5};
    \wirelabel{\footnotesize 2}{19.30451127819549,4};
    \wirelabel{\footnotesize 4}{19.30451127819549,3};
    \wirelabel{\footnotesize 7}{19.30451127819549,2};
    \wirelabel{\footnotesize 1}{19.30451127819549,1};
    \wirelabel{\footnotesize 5}{19.30451127819549,0};
    \biggate{$T$}{20.770676691729335,5}{20.770676691729335,7};
    \biggate{$T$}{20.770676691729335,2}{20.770676691729335,4};
    \dotswapgate{21.992481203007515}{6}{5};
    \dotswapgate{21.992481203007515}{3}{2};
    \dotswapgate{22.969924812030087}{5}{4};
    \dotswapgate{22.969924812030087}{2}{1};
    \dotswapgate{23.94736842105263}{4}{3};
    \dotswapgate{23.94736842105263}{1}{0};
    \dotswapgate{24.924812030075202}{3}{2};
    \dotswapgate{25.902255639097746}{4}{3};
    \dotswapgate{25.902255639097746}{2}{1};
    \dotswapgate{26.879699248120318}{5}{4};
    \dotswapgate{26.879699248120318}{3}{2};
    \dotswapgate{27.85714285714286}{6}{5};
    \wirelabel{\footnotesize 0}{29.07894736842107,7};
    \wirelabel{\footnotesize 1}{29.07894736842107,6};
    \wirelabel{\footnotesize 6}{29.07894736842107,5};
    \wirelabel{\footnotesize 2}{29.07894736842107,4};
    \wirelabel{\footnotesize 5}{29.07894736842107,3};
    \wirelabel{\footnotesize 7}{29.07894736842107,2};
    \wirelabel{\footnotesize 3}{29.07894736842107,1};
    \wirelabel{\footnotesize 4}{29.07894736842107,0};
    \biggate{$T$}{30.545112781954884,5}{30.545112781954884,7};
    \biggate{$T$}{30.545112781954884,2}{30.545112781954884,4};
    \dotswapgate{31.766917293233092}{6}{5};
    \dotswapgate{31.766917293233092}{3}{2};
    \dotswapgate{32.744360902255636}{7}{6};
    \dotswapgate{32.744360902255636}{5}{4};
    \dotswapgate{32.744360902255636}{2}{1};
    \dotswapgate{33.72180451127821}{6}{5};
    \dotswapgate{33.72180451127821}{4}{3};
    \dotswapgate{34.69924812030075}{7}{6};
    \dotswapgate{34.69924812030075}{5}{4};
    \dotswapgate{34.69924812030075}{3}{2};
    \dotswapgate{35.67669172932332}{4}{3};
    \dotswapgate{35.67669172932332}{2}{1};
    \dotswapgate{36.65413533834587}{3}{2};
    \dotswapgate{36.65413533834587}{1}{0};
    \dotswapgate{37.63157894736841}{2}{1};
    \wirelabel{\footnotesize 2}{38.85338345864662,7};
    \wirelabel{\footnotesize 6}{38.85338345864662,6};
    \wirelabel{\footnotesize 7}{38.85338345864662,5};
    \wirelabel{\footnotesize 3}{38.85338345864662,4};
    \wirelabel{\footnotesize 5}{38.85338345864662,3};
    \wirelabel{\footnotesize 4}{38.85338345864662,2};
    \wirelabel{\footnotesize 0}{38.85338345864662,1};
    \wirelabel{\footnotesize 1}{38.85338345864662,0};
    \biggate{$T$}{40.31954887218046,5}{40.31954887218046,7};
    \biggate{$T$}{40.31954887218046,2}{40.31954887218046,4};
    \dotswapgate{41.54135338345864}{6}{5};
    \dotswapgate{41.54135338345864}{4}{3};
    \dotswapgate{41.54135338345864}{2}{1};
    \dotswapgate{42.518796992481214}{5}{4};
    \dotswapgate{42.518796992481214}{3}{2};
    \dotswapgate{43.49624060150376}{6}{5};
    \dotswapgate{43.49624060150376}{4}{3};
    \dotswapgate{44.47368421052633}{7}{6};
    \dotswapgate{44.47368421052633}{5}{4};
    \dotswapgate{44.47368421052633}{3}{2};
    \dotswapgate{45.45112781954887}{6}{5};
    \dotswapgate{45.45112781954887}{4}{3};
    \dotswapgate{45.45112781954887}{2}{1};
    \dotswapgate{46.428571428571416}{7}{6};
    \dotswapgate{46.428571428571416}{3}{2};
    \dotswapgate{46.428571428571416}{1}{0};
    \wirelabel{\footnotesize 0}{47.650375939849624,7};
    \wirelabel{\footnotesize 5}{47.650375939849624,6};
    \wirelabel{\footnotesize 2}{47.650375939849624,5};
    \wirelabel{\footnotesize 3}{47.650375939849624,4};
    \wirelabel{\footnotesize 4}{47.650375939849624,3};
    \wirelabel{\footnotesize 7}{47.650375939849624,2};
    \wirelabel{\footnotesize 1}{47.650375939849624,1};
    \wirelabel{\footnotesize 6}{47.650375939849624,0};
    \biggate{$T$}{49.11654135338347,5}{49.11654135338347,7};
    \biggate{$T$}{49.11654135338347,2}{49.11654135338347,4};
    \dotswapgate{50.33834586466165}{5}{4};
    \dotswapgate{50.33834586466165}{2}{1};
    \dotswapgate{51.31578947368422}{6}{5};
    \dotswapgate{51.31578947368422}{4}{3};
    \dotswapgate{52.29323308270676}{5}{4};
    \dotswapgate{52.29323308270676}{3}{2};
    \dotswapgate{53.270676691729335}{6}{5};
    \dotswapgate{53.270676691729335}{4}{3};
    \dotswapgate{53.270676691729335}{2}{1};
    \dotswapgate{54.24812030075188}{5}{4};
    \dotswapgate{55.22556390977445}{6}{5};
    \wirelabel{\footnotesize 0}{56.44736842105263,7};
    \wirelabel{\footnotesize 1}{56.44736842105263,6};
    \wirelabel{\footnotesize 4}{56.44736842105263,5};
    \wirelabel{\footnotesize 3}{56.44736842105263,4};
    \wirelabel{\footnotesize 5}{56.44736842105263,3};
    \wirelabel{\footnotesize 7}{56.44736842105263,2};
    \wirelabel{\footnotesize 2}{56.44736842105263,1};
    \wirelabel{\footnotesize 6}{56.44736842105263,0};
    \biggate{$T$}{57.913533834586474,5}{57.913533834586474,7};
    \biggate{$T$}{57.913533834586474,2}{57.913533834586474,4};
    \dotswapgate{59.13533834586465}{5}{4};
    \dotswapgate{59.13533834586465}{3}{2};
    \dotswapgate{60.112781954887225}{4}{3};
    \dotswapgate{60.112781954887225}{2}{1};
    \dotswapgate{61.09022556390977}{3}{2};
    \dotswapgate{61.09022556390977}{1}{0};
    \dotswapgate{62.06766917293234}{4}{3};
    \dotswapgate{62.06766917293234}{2}{1};
    \dotswapgate{63.045112781954884}{5}{4};
    \dotswapgate{63.045112781954884}{3}{2};
    \wirelabel{\footnotesize 0}{64.26691729323309,7};
    \wirelabel{\footnotesize 1}{64.26691729323309,6};
    \wirelabel{\footnotesize 2}{64.26691729323309,5};
    \wirelabel{\footnotesize 3}{64.26691729323309,4};
    \wirelabel{\footnotesize 6}{64.26691729323309,3};
    \wirelabel{\footnotesize 7}{64.26691729323309,2};
    \wirelabel{\footnotesize 4}{64.26691729323309,1};
    \wirelabel{\footnotesize 5}{64.26691729323309,0};
  \end{qcircuit}
}
\]
\[
  \scalebox{0.45}{
  \begin{qcircuit}[scale=0.4]
    \grid{65.0}{0,1,2,3,4,5,6,7}
    \wirelabel{\footnotesize 0}{0.7442748091603164,7};
    \wirelabel{\footnotesize 1}{0.7442748091603164,6};
    \wirelabel{\footnotesize 2}{0.7442748091603164,5};
    \wirelabel{\footnotesize 3}{0.7442748091603164,4};
    \wirelabel{\footnotesize 6}{0.7442748091603164,3};
    \wirelabel{\footnotesize 7}{0.7442748091603164,2};
    \wirelabel{\footnotesize 4}{0.7442748091603164,1};
    \wirelabel{\footnotesize 5}{0.7442748091603164,0};
    \biggate{$T$}{2.232824427480921,5}{2.232824427480921,7};
    \biggate{$T$}{2.232824427480921,2}{2.232824427480921,4};
    \dotswapgate{3.4732824427480864}{7}{6};
    \dotswapgate{3.4732824427480864}{5}{4};
    \dotswapgate{3.4732824427480864}{2}{1};
    \dotswapgate{4.465648854961842}{6}{5};
    \dotswapgate{4.465648854961842}{4}{3};
    \dotswapgate{4.465648854961842}{1}{0};
    \dotswapgate{5.458015267175568}{5}{4};
    \dotswapgate{5.458015267175568}{3}{2};
    \dotswapgate{6.450381679389324}{6}{5};
    \dotswapgate{6.450381679389324}{4}{3};
    \dotswapgate{6.450381679389324}{2}{1};
    \dotswapgate{7.44274809160305}{3}{2};
    \dotswapgate{7.44274809160305}{1}{0};
    \dotswapgate{8.435114503816806}{2}{1};
    \wirelabel{\footnotesize 1}{9.675572519083971,7};
    \wirelabel{\footnotesize 6}{9.675572519083971,6};
    \wirelabel{\footnotesize 3}{9.675572519083971,5};
    \wirelabel{\footnotesize 4}{9.675572519083971,4};
    \wirelabel{\footnotesize 5}{9.675572519083971,3};
    \wirelabel{\footnotesize 7}{9.675572519083971,2};
    \wirelabel{\footnotesize 0}{9.675572519083971,1};
    \wirelabel{\footnotesize 2}{9.675572519083971,0};
    \biggate{$T$}{11.164122137404576,5}{11.164122137404576,7};
    \biggate{$T$}{11.164122137404576,2}{11.164122137404576,4};
    \dotswapgate{12.40458015267177}{5}{4};
    \dotswapgate{12.40458015267177}{1}{0};
    \dotswapgate{13.396946564885496}{6}{5};
    \dotswapgate{13.396946564885496}{4}{3};
    \dotswapgate{13.396946564885496}{2}{1};
    \dotswapgate{14.389312977099252}{5}{4};
    \dotswapgate{14.389312977099252}{3}{2};
    \dotswapgate{15.381679389312978}{6}{5};
    \dotswapgate{15.381679389312978}{4}{3};
    \dotswapgate{15.381679389312978}{2}{1};
    \dotswapgate{16.374045801526734}{5}{4};
    \dotswapgate{16.374045801526734}{1}{0};
    \dotswapgate{17.36641221374046}{6}{5};
    \wirelabel{\footnotesize 1}{18.606870229007654,7};
    \wirelabel{\footnotesize 2}{18.606870229007654,6};
    \wirelabel{\footnotesize 5}{18.606870229007654,5};
    \wirelabel{\footnotesize 4}{18.606870229007654,4};
    \wirelabel{\footnotesize 6}{18.606870229007654,3};
    \wirelabel{\footnotesize 7}{18.606870229007654,2};
    \wirelabel{\footnotesize 0}{18.606870229007654,1};
    \wirelabel{\footnotesize 3}{18.606870229007654,0};
    \biggate{$T$}{20.09541984732826,5}{20.09541984732826,7};
    \biggate{$T$}{20.09541984732826,2}{20.09541984732826,4};
    \dotswapgate{21.335877862595424}{7}{6};
    \dotswapgate{21.335877862595424}{5}{4};
    \dotswapgate{21.335877862595424}{2}{1};
    \dotswapgate{22.32824427480918}{6}{5};
    \dotswapgate{22.32824427480918}{3}{2};
    \dotswapgate{23.320610687022906}{7}{6};
    \dotswapgate{23.320610687022906}{5}{4};
    \dotswapgate{24.31297709923666}{4}{3};
    \dotswapgate{25.30534351145039}{5}{4};
    \dotswapgate{25.30534351145039}{3}{2};
    \dotswapgate{26.297709923664144}{6}{5};
    \dotswapgate{26.297709923664144}{2}{1};
    \dotswapgate{27.29007633587787}{7}{6};
    \wirelabel{\footnotesize 0}{28.530534351145036,7};
    \wirelabel{\footnotesize 4}{28.530534351145036,6};
    \wirelabel{\footnotesize 2}{28.530534351145036,5};
    \wirelabel{\footnotesize 5}{28.530534351145036,4};
    \wirelabel{\footnotesize 6}{28.530534351145036,3};
    \wirelabel{\footnotesize 7}{28.530534351145036,2};
    \wirelabel{\footnotesize 1}{28.530534351145036,1};
    \wirelabel{\footnotesize 3}{28.530534351145036,0};
    \biggate{$T$}{30.01908396946567,5}{30.01908396946567,7};
    \biggate{$T$}{30.01908396946567,2}{30.01908396946567,4};
    \dotswapgate{31.259541984732834}{7}{6};
    \dotswapgate{31.259541984732834}{2}{1};
    \dotswapgate{32.25190839694656}{6}{5};
    \dotswapgate{32.25190839694656}{3}{2};
    \dotswapgate{32.25190839694656}{1}{0};
    \dotswapgate{33.244274809160316}{5}{4};
    \dotswapgate{33.244274809160316}{2}{1};
    \dotswapgate{34.23664122137404}{4}{3};
    \dotswapgate{35.2290076335878}{5}{4};
    \dotswapgate{35.2290076335878}{3}{2};
    \dotswapgate{36.221374045801525}{6}{5};
    \dotswapgate{36.221374045801525}{4}{3};
    \dotswapgate{36.221374045801525}{2}{1};
    \dotswapgate{37.21374045801528}{7}{6};
    \dotswapgate{37.21374045801528}{1}{0};
    \wirelabel{\footnotesize 1}{38.454198473282446,7};
    \wirelabel{\footnotesize 4}{38.454198473282446,6};
    \wirelabel{\footnotesize 2}{38.454198473282446,5};
    \wirelabel{\footnotesize 3}{38.454198473282446,4};
    \wirelabel{\footnotesize 5}{38.454198473282446,3};
    \wirelabel{\footnotesize 6}{38.454198473282446,2};
    \wirelabel{\footnotesize 7}{38.454198473282446,1};
    \wirelabel{\footnotesize 0}{38.454198473282446,0};
    \biggate{$T$}{39.94274809160305,5}{39.94274809160305,7};
    \biggate{$T$}{39.94274809160305,2}{39.94274809160305,4};
    \dotswapgate{41.183206106870244}{5}{4};
    \dotswapgate{41.183206106870244}{1}{0};
    \dotswapgate{42.17557251908397}{6}{5};
    \dotswapgate{42.17557251908397}{4}{3};
    \dotswapgate{42.17557251908397}{2}{1};
    \dotswapgate{43.167938931297726}{5}{4};
    \dotswapgate{43.167938931297726}{3}{2};
    \dotswapgate{44.16030534351145}{6}{5};
    \dotswapgate{44.16030534351145}{4}{3};
    \dotswapgate{44.16030534351145}{2}{1};
    \dotswapgate{45.15267175572521}{5}{4};
    \dotswapgate{45.15267175572521}{3}{2};
    \dotswapgate{45.15267175572521}{1}{0};
    \dotswapgate{46.145038167938935}{6}{5};
    \dotswapgate{47.13740458015269}{7}{6};
    \wirelabel{\footnotesize 0}{48.377862595419856,7};
    \wirelabel{\footnotesize 1}{48.377862595419856,6};
    \wirelabel{\footnotesize 5}{48.377862595419856,5};
    \wirelabel{\footnotesize 3}{48.377862595419856,4};
    \wirelabel{\footnotesize 6}{48.377862595419856,3};
    \wirelabel{\footnotesize 4}{48.377862595419856,2};
    \wirelabel{\footnotesize 7}{48.377862595419856,1};
    \wirelabel{\footnotesize 2}{48.377862595419856,0};
    \biggate{$T$}{49.86641221374046,5}{49.86641221374046,7};
    \biggate{$T$}{49.86641221374046,2}{49.86641221374046,4};
    \dotswapgate{51.106870229007654}{5}{4};
    \dotswapgate{51.106870229007654}{2}{1};
    \dotswapgate{52.09923664122138}{6}{5};
    \dotswapgate{52.09923664122138}{4}{3};
    \dotswapgate{52.09923664122138}{1}{0};
    \dotswapgate{53.09160305343511}{2}{1};
    \dotswapgate{54.08396946564886}{3}{2};
    \dotswapgate{55.07633587786259}{4}{3};
    \dotswapgate{56.068702290076345}{5}{4};
    \dotswapgate{57.06106870229007}{6}{5};
    \wirelabel{\footnotesize 0}{58.301526717557266,7};
    \wirelabel{\footnotesize 2}{58.301526717557266,6};
    \wirelabel{\footnotesize 3}{58.301526717557266,5};
    \wirelabel{\footnotesize 1}{58.301526717557266,4};
    \wirelabel{\footnotesize 6}{58.301526717557266,3};
    \wirelabel{\footnotesize 5}{58.301526717557266,2};
    \wirelabel{\footnotesize 7}{58.301526717557266,1};
    \wirelabel{\footnotesize 4}{58.301526717557266,0};
    \biggate{$T$}{59.79007633587787,5}{59.79007633587787,7};
    \biggate{$T$}{59.79007633587787,2}{59.79007633587787,4};
    \dotswapgate{61.030534351145036}{6}{5};
    \dotswapgate{61.030534351145036}{3}{2};
    \dotswapgate{61.030534351145036}{1}{0};
    \dotswapgate{62.02290076335879}{5}{4};
    \dotswapgate{62.02290076335879}{2}{1};
    \dotswapgate{63.01526717557252}{3}{2};
    \dotswapgate{63.01526717557252}{1}{0};
    \wirelabel{\footnotesize 0}{64.25572519083971,7};
    \wirelabel{\footnotesize 3}{64.25572519083971,6};
    \wirelabel{\footnotesize 1}{64.25572519083971,5};
    \wirelabel{\footnotesize 2}{64.25572519083971,4};
    \wirelabel{\footnotesize 4}{64.25572519083971,3};
    \wirelabel{\footnotesize 5}{64.25572519083971,2};
    \wirelabel{\footnotesize 7}{64.25572519083971,1};
    \wirelabel{\footnotesize 6}{64.25572519083971,0};
  \end{qcircuit}
}
\]
\[
  \scalebox{0.45}{
  \begin{qcircuit}[scale=0.4]
    \grid{65.0}{0,1,2,3,4,5,6,7}
    \wirelabel{\footnotesize 0}{0.7442748091603164,7};
    \wirelabel{\footnotesize 3}{0.7442748091603164,6};
    \wirelabel{\footnotesize 1}{0.7442748091603164,5};
    \wirelabel{\footnotesize 2}{0.7442748091603164,4};
    \wirelabel{\footnotesize 4}{0.7442748091603164,3};
    \wirelabel{\footnotesize 5}{0.7442748091603164,2};
    \wirelabel{\footnotesize 7}{0.7442748091603164,1};
    \wirelabel{\footnotesize 6}{0.7442748091603164,0};
    \biggate{$T$}{2.232824427480921,5}{2.232824427480921,7};
    \biggate{$T$}{2.232824427480921,2}{2.232824427480921,4};
    \dotswapgate{3.4732824427480864}{5}{4};
    \dotswapgate{3.4732824427480864}{1}{0};
    \dotswapgate{4.465648854961842}{6}{5};
    \dotswapgate{4.465648854961842}{4}{3};
    \dotswapgate{4.465648854961842}{2}{1};
    \dotswapgate{5.458015267175597}{5}{4};
    \dotswapgate{5.458015267175597}{3}{2};
    \dotswapgate{6.450381679389295}{6}{5};
    \dotswapgate{6.450381679389295}{4}{3};
    \dotswapgate{6.450381679389295}{2}{1};
    \dotswapgate{7.44274809160305}{5}{4};
    \dotswapgate{7.44274809160305}{3}{2};
    \dotswapgate{7.44274809160305}{1}{0};
    \wirelabel{\footnotesize 0}{8.683206106870216,7};
    \wirelabel{\footnotesize 4}{8.683206106870216,6};
    \wirelabel{\footnotesize 6}{8.683206106870216,5};
    \wirelabel{\footnotesize 2}{8.683206106870216,4};
    \wirelabel{\footnotesize 5}{8.683206106870216,3};
    \wirelabel{\footnotesize 3}{8.683206106870216,2};
    \wirelabel{\footnotesize 7}{8.683206106870216,1};
    \wirelabel{\footnotesize 1}{8.683206106870216,0};
    \biggate{$T$}{10.171755725190849,5}{10.171755725190849,7};
    \biggate{$T$}{10.171755725190849,2}{10.171755725190849,4};
    \dotswapgate{11.412213740458014}{5}{4};
    \dotswapgate{11.412213740458014}{2}{1};
    \dotswapgate{12.40458015267177}{4}{3};
    \dotswapgate{12.40458015267177}{1}{0};
    \dotswapgate{13.396946564885525}{5}{4};
    \dotswapgate{13.396946564885525}{2}{1};
    \dotswapgate{14.389312977099223}{6}{5};
    \dotswapgate{14.389312977099223}{3}{2};
    \dotswapgate{15.381679389312978}{4}{3};
    \wirelabel{\footnotesize 0}{16.622137404580144,7};
    \wirelabel{\footnotesize 5}{16.622137404580144,6};
    \wirelabel{\footnotesize 4}{16.622137404580144,5};
    \wirelabel{\footnotesize 1}{16.622137404580144,4};
    \wirelabel{\footnotesize 2}{16.622137404580144,3};
    \wirelabel{\footnotesize 6}{16.622137404580144,2};
    \wirelabel{\footnotesize 7}{16.622137404580144,1};
    \wirelabel{\footnotesize 3}{16.622137404580144,0};
    \biggate{$T$}{18.110687022900777,5}{18.110687022900777,7};
    \biggate{$T$}{18.110687022900777,2}{18.110687022900777,4};
    \dotswapgate{19.351145038167942}{6}{5};
    \dotswapgate{19.351145038167942}{4}{3};
    \dotswapgate{19.351145038167942}{1}{0};
    \dotswapgate{20.343511450381698}{5}{4};
    \dotswapgate{20.343511450381698}{3}{2};
    \dotswapgate{21.335877862595396}{6}{5};
    \dotswapgate{21.335877862595396}{4}{3};
    \dotswapgate{22.32824427480915}{5}{4};
    \dotswapgate{22.32824427480915}{3}{2};
    \dotswapgate{23.320610687022906}{6}{5};
    \dotswapgate{23.320610687022906}{4}{3};
    \dotswapgate{23.320610687022906}{2}{1};
    \dotswapgate{24.31297709923666}{3}{2};
    \dotswapgate{24.31297709923666}{1}{0};
    \wirelabel{\footnotesize 0}{25.553435114503827,7};
    \wirelabel{\footnotesize 6}{25.553435114503827,6};
    \wirelabel{\footnotesize 2}{25.553435114503827,5};
    \wirelabel{\footnotesize 1}{25.553435114503827,4};
    \wirelabel{\footnotesize 3}{25.553435114503827,3};
    \wirelabel{\footnotesize 4}{25.553435114503827,2};
    \wirelabel{\footnotesize 7}{25.553435114503827,1};
    \wirelabel{\footnotesize 5}{25.553435114503827,0};
    \biggate{$T$}{27.041984732824403,5}{27.041984732824403,7};
    \biggate{$T$}{27.041984732824403,2}{27.041984732824403,4};
    \dotswapgate{28.282442748091626}{6}{5};
    \dotswapgate{28.282442748091626}{2}{1};
    \dotswapgate{29.274809160305324}{5}{4};
    \dotswapgate{29.274809160305324}{3}{2};
    \dotswapgate{29.274809160305324}{1}{0};
    \dotswapgate{30.26717557251908}{6}{5};
    \dotswapgate{30.26717557251908}{4}{3};
    \dotswapgate{31.259541984732834}{5}{4};
    \dotswapgate{31.259541984732834}{3}{2};
    \dotswapgate{32.25190839694659}{4}{3};
    \dotswapgate{32.25190839694659}{2}{1};
    \dotswapgate{33.24427480916029}{1}{0};
    \wirelabel{\footnotesize 0}{34.484732824427454,7};
    \wirelabel{\footnotesize 1}{34.484732824427454,6};
    \wirelabel{\footnotesize 7}{34.484732824427454,5};
    \wirelabel{\footnotesize 3}{34.484732824427454,4};
    \wirelabel{\footnotesize 2}{34.484732824427454,3};
    \wirelabel{\footnotesize 5}{34.484732824427454,2};
    \wirelabel{\footnotesize 4}{34.484732824427454,1};
    \wirelabel{\footnotesize 6}{34.484732824427454,0};
    \biggate{$Q$}{35.973282442748086,4}{35.973282442748086,7};
    \biggate{$Q$}{35.973282442748086,0}{35.973282442748086,3};
    \dotswapgate{37.21374045801525}{6}{5};
    \dotswapgate{37.21374045801525}{4}{3};
    \dotswapgate{38.20610687022901}{7}{6};
    \dotswapgate{38.20610687022901}{5}{4};
    \dotswapgate{38.20610687022901}{3}{2};
    \dotswapgate{39.19847328244276}{6}{5};
    \dotswapgate{39.19847328244276}{4}{3};
    \dotswapgate{40.19083969465646}{7}{6};
    \dotswapgate{40.19083969465646}{5}{4};
    \dotswapgate{40.19083969465646}{3}{2};
    \wirelabel{\footnotesize 2}{41.43129770992368,7};
    \wirelabel{\footnotesize 7}{41.43129770992368,6};
    \wirelabel{\footnotesize 5}{41.43129770992368,5};
    \wirelabel{\footnotesize 0}{41.43129770992368,4};
    \wirelabel{\footnotesize 3}{41.43129770992368,3};
    \wirelabel{\footnotesize 1}{41.43129770992368,2};
    \wirelabel{\footnotesize 4}{41.43129770992368,1};
    \wirelabel{\footnotesize 6}{41.43129770992368,0};
    \biggate{$Q$}{42.91984732824426,4}{42.91984732824426,7};
    \biggate{$Q$}{42.91984732824426,0}{42.91984732824426,3};
    \dotswapgate{44.160305343511425}{6}{5};
    \dotswapgate{44.160305343511425}{4}{3};
    \dotswapgate{45.15267175572518}{7}{6};
    \dotswapgate{45.15267175572518}{5}{4};
    \dotswapgate{45.15267175572518}{3}{2};
    \dotswapgate{46.145038167938935}{6}{5};
    \dotswapgate{46.145038167938935}{4}{3};
    \dotswapgate{47.13740458015269}{7}{6};
    \dotswapgate{47.13740458015269}{5}{4};
    \dotswapgate{47.13740458015269}{3}{2};
    \wirelabel{\footnotesize 3}{48.377862595419856,7};
    \wirelabel{\footnotesize 5}{48.377862595419856,6};
    \wirelabel{\footnotesize 1}{48.377862595419856,5};
    \wirelabel{\footnotesize 2}{48.377862595419856,4};
    \wirelabel{\footnotesize 0}{48.377862595419856,3};
    \wirelabel{\footnotesize 7}{48.377862595419856,2};
    \wirelabel{\footnotesize 4}{48.377862595419856,1};
    \wirelabel{\footnotesize 6}{48.377862595419856,0};
    \biggate{$Q$}{49.86641221374043,4}{49.86641221374043,7};
    \biggate{$Q$}{49.86641221374043,0}{49.86641221374043,3};
    \dotswapgate{51.106870229007654}{7}{6};
    \dotswapgate{51.106870229007654}{4}{3};
    \dotswapgate{51.106870229007654}{1}{0};
    \dotswapgate{52.09923664122135}{6}{5};
    \dotswapgate{52.09923664122135}{3}{2};
    \dotswapgate{53.09160305343511}{7}{6};
    \dotswapgate{53.09160305343511}{5}{4};
    \dotswapgate{54.08396946564886}{6}{5};
    \dotswapgate{54.08396946564886}{4}{3};
    \dotswapgate{55.07633587786262}{7}{6};
    \dotswapgate{55.07633587786262}{5}{4};
    \dotswapgate{55.07633587786262}{3}{2};
    \dotswapgate{56.06870229007632}{2}{1};
    \wirelabel{\footnotesize 0}{57.30916030534354,7};
    \wirelabel{\footnotesize 1}{57.30916030534354,6};
    \wirelabel{\footnotesize 7}{57.30916030534354,5};
    \wirelabel{\footnotesize 5}{57.30916030534354,4};
    \wirelabel{\footnotesize 2}{57.30916030534354,3};
    \wirelabel{\footnotesize 6}{57.30916030534354,2};
    \wirelabel{\footnotesize 3}{57.30916030534354,1};
    \wirelabel{\footnotesize 4}{57.30916030534354,0};
    \biggate{$Q$}{58.797709923664115,4}{58.797709923664115,7};
    \biggate{$Q$}{58.797709923664115,0}{58.797709923664115,3};
    \dotswapgate{60.03816793893128}{6}{5};
    \dotswapgate{60.03816793893128}{4}{3};
    \dotswapgate{61.030534351145036}{7}{6};
    \dotswapgate{61.030534351145036}{5}{4};
    \dotswapgate{61.030534351145036}{3}{2};
    \dotswapgate{62.02290076335879}{6}{5};
    \dotswapgate{62.02290076335879}{4}{3};
    \dotswapgate{63.015267175572546}{7}{6};
    \dotswapgate{63.015267175572546}{5}{4};
    \dotswapgate{63.015267175572546}{3}{2};
    \wirelabel{\footnotesize 2}{64.25572519083971,7};
    \wirelabel{\footnotesize 7}{64.25572519083971,6};
    \wirelabel{\footnotesize 6}{64.25572519083971,5};
    \wirelabel{\footnotesize 0}{64.25572519083971,4};
    \wirelabel{\footnotesize 5}{64.25572519083971,3};
    \wirelabel{\footnotesize 1}{64.25572519083971,2};
    \wirelabel{\footnotesize 3}{64.25572519083971,1};
    \wirelabel{\footnotesize 4}{64.25572519083971,0};
  \end{qcircuit}
}
\]
\[
  \scalebox{0.45}{
  \begin{qcircuit}[scale=0.4]
    \grid{65.0}{0,1,2,3,4,5,6,7}
    \wirelabel{\footnotesize 2}{0.7330827067669361,7};
    \wirelabel{\footnotesize 7}{0.7330827067669361,6};
    \wirelabel{\footnotesize 6}{0.7330827067669361,5};
    \wirelabel{\footnotesize 0}{0.7330827067669361,4};
    \wirelabel{\footnotesize 5}{0.7330827067669361,3};
    \wirelabel{\footnotesize 1}{0.7330827067669361,2};
    \wirelabel{\footnotesize 3}{0.7330827067669361,1};
    \wirelabel{\footnotesize 4}{0.7330827067669361,0};
    \biggate{$Q$}{2.1992481203007515,4}{2.1992481203007515,7};
    \biggate{$Q$}{2.1992481203007515,0}{2.1992481203007515,3};
    \dotswapgate{3.4210526315789593}{6}{5};
    \dotswapgate{3.4210526315789593}{4}{3};
    \dotswapgate{4.398496240601503}{7}{6};
    \dotswapgate{4.398496240601503}{5}{4};
    \dotswapgate{4.398496240601503}{3}{2};
    \dotswapgate{5.375939849624103}{6}{5};
    \dotswapgate{5.375939849624103}{4}{3};
    \dotswapgate{6.353383458646647}{7}{6};
    \dotswapgate{6.353383458646647}{5}{4};
    \dotswapgate{6.353383458646647}{3}{2};
    \wirelabel{\footnotesize 5}{7.575187969924855,7};
    \wirelabel{\footnotesize 6}{7.575187969924855,6};
    \wirelabel{\footnotesize 1}{7.575187969924855,5};
    \wirelabel{\footnotesize 2}{7.575187969924855,4};
    \wirelabel{\footnotesize 0}{7.575187969924855,3};
    \wirelabel{\footnotesize 7}{7.575187969924855,2};
    \wirelabel{\footnotesize 3}{7.575187969924855,1};
    \wirelabel{\footnotesize 4}{7.575187969924855,0};
    \biggate{$Q$}{9.04135338345867,4}{9.04135338345867,7};
    \biggate{$Q$}{9.04135338345867,0}{9.04135338345867,3};
    \dotswapgate{10.263157894736878}{7}{6};
    \dotswapgate{10.263157894736878}{4}{3};
    \dotswapgate{11.240601503759422}{6}{5};
    \dotswapgate{11.240601503759422}{3}{2};
    \dotswapgate{12.218045112781965}{7}{6};
    \dotswapgate{12.218045112781965}{5}{4};
    \dotswapgate{13.195488721804509}{6}{5};
    \dotswapgate{13.195488721804509}{4}{3};
    \dotswapgate{14.17293233082711}{7}{6};
    \dotswapgate{14.17293233082711}{5}{4};
    \dotswapgate{14.17293233082711}{3}{2};
    \dotswapgate{15.150375939849653}{2}{1};
    \dotswapgate{16.127819548872196}{1}{0};
    \wirelabel{\footnotesize 0}{17.349624060150404,7};
    \wirelabel{\footnotesize 1}{17.349624060150404,6};
    \wirelabel{\footnotesize 7}{17.349624060150404,5};
    \wirelabel{\footnotesize 6}{17.349624060150404,4};
    \wirelabel{\footnotesize 2}{17.349624060150404,3};
    \wirelabel{\footnotesize 3}{17.349624060150404,2};
    \wirelabel{\footnotesize 4}{17.349624060150404,1};
    \wirelabel{\footnotesize 5}{17.349624060150404,0};
    \biggate{$Q$}{18.81578947368422,4}{18.81578947368422,7};
    \biggate{$Q$}{18.81578947368422,0}{18.81578947368422,3};
    \dotswapgate{20.037593984962427}{6}{5};
    \dotswapgate{20.037593984962427}{4}{3};
    \dotswapgate{21.01503759398497}{7}{6};
    \dotswapgate{21.01503759398497}{5}{4};
    \dotswapgate{21.01503759398497}{3}{2};
    \dotswapgate{21.992481203007515}{6}{5};
    \dotswapgate{21.992481203007515}{4}{3};
    \dotswapgate{22.969924812030115}{7}{6};
    \dotswapgate{22.969924812030115}{5}{4};
    \dotswapgate{22.969924812030115}{3}{2};
    \wirelabel{\footnotesize 2}{24.191729323308266,7};
    \wirelabel{\footnotesize 7}{24.191729323308266,6};
    \wirelabel{\footnotesize 3}{24.191729323308266,5};
    \wirelabel{\footnotesize 0}{24.191729323308266,4};
    \wirelabel{\footnotesize 6}{24.191729323308266,3};
    \wirelabel{\footnotesize 1}{24.191729323308266,2};
    \wirelabel{\footnotesize 4}{24.191729323308266,1};
    \wirelabel{\footnotesize 5}{24.191729323308266,0};
    \biggate{$Q$}{25.657894736842138,4}{25.657894736842138,7};
    \biggate{$Q$}{25.657894736842138,0}{25.657894736842138,3};
    \dotswapgate{26.879699248120346}{6}{5};
    \dotswapgate{26.879699248120346}{4}{3};
    \dotswapgate{27.85714285714289}{7}{6};
    \dotswapgate{27.85714285714289}{5}{4};
    \dotswapgate{27.85714285714289}{3}{2};
    \dotswapgate{28.834586466165433}{6}{5};
    \dotswapgate{28.834586466165433}{4}{3};
    \dotswapgate{29.812030075187977}{7}{6};
    \dotswapgate{29.812030075187977}{5}{4};
    \dotswapgate{29.812030075187977}{3}{2};
    \wirelabel{\footnotesize 6}{31.033834586466185,7};
    \wirelabel{\footnotesize 3}{31.033834586466185,6};
    \wirelabel{\footnotesize 1}{31.033834586466185,5};
    \wirelabel{\footnotesize 2}{31.033834586466185,4};
    \wirelabel{\footnotesize 0}{31.033834586466185,3};
    \wirelabel{\footnotesize 7}{31.033834586466185,2};
    \wirelabel{\footnotesize 4}{31.033834586466185,1};
    \wirelabel{\footnotesize 5}{31.033834586466185,0};
    \biggate{$Q$}{32.5,4}{32.5,7};
    \biggate{$Q$}{32.5,0}{32.5,3};
    \dotswapgate{33.72180451127821}{7}{6};
    \dotswapgate{33.72180451127821}{4}{3};
    \dotswapgate{34.69924812030075}{6}{5};
    \dotswapgate{35.67669172932335}{7}{6};
    \dotswapgate{35.67669172932335}{5}{4};
    \dotswapgate{36.654135338345895}{6}{5};
    \dotswapgate{36.654135338345895}{4}{3};
    \dotswapgate{37.63157894736844}{7}{6};
    \dotswapgate{37.63157894736844}{5}{4};
    \dotswapgate{37.63157894736844}{3}{2};
    \dotswapgate{38.60902255639098}{4}{3};
    \dotswapgate{38.60902255639098}{2}{1};
    \dotswapgate{39.58646616541358}{1}{0};
    \wirelabel{\footnotesize 0}{40.808270676691734,7};
    \wirelabel{\footnotesize 1}{40.808270676691734,6};
    \wirelabel{\footnotesize 2}{40.808270676691734,5};
    \wirelabel{\footnotesize 7}{40.808270676691734,4};
    \wirelabel{\footnotesize 3}{40.808270676691734,3};
    \wirelabel{\footnotesize 4}{40.808270676691734,2};
    \wirelabel{\footnotesize 5}{40.808270676691734,1};
    \wirelabel{\footnotesize 6}{40.808270676691734,0};
    \biggate{$Q$}{42.274436090225606,4}{42.274436090225606,7};
    \biggate{$Q$}{42.274436090225606,0}{42.274436090225606,3};
    \dotswapgate{43.49624060150376}{6}{5};
    \dotswapgate{43.49624060150376}{4}{3};
    \dotswapgate{44.47368421052636}{7}{6};
    \dotswapgate{44.47368421052636}{5}{4};
    \dotswapgate{44.47368421052636}{3}{2};
    \dotswapgate{45.4511278195489}{6}{5};
    \dotswapgate{45.4511278195489}{4}{3};
    \dotswapgate{46.428571428571445}{7}{6};
    \dotswapgate{46.428571428571445}{5}{4};
    \dotswapgate{46.428571428571445}{3}{2};
    \wirelabel{\footnotesize 3}{47.65037593984965,7};
    \wirelabel{\footnotesize 2}{47.65037593984965,6};
    \wirelabel{\footnotesize 4}{47.65037593984965,5};
    \wirelabel{\footnotesize 0}{47.65037593984965,4};
    \wirelabel{\footnotesize 7}{47.65037593984965,3};
    \wirelabel{\footnotesize 1}{47.65037593984965,2};
    \wirelabel{\footnotesize 5}{47.65037593984965,1};
    \wirelabel{\footnotesize 6}{47.65037593984965,0};
    \biggate{$Q$}{49.11654135338347,4}{49.11654135338347,7};
    \biggate{$Q$}{49.11654135338347,0}{49.11654135338347,3};
    \dotswapgate{50.338345864661676}{6}{5};
    \dotswapgate{50.338345864661676}{4}{3};
    \dotswapgate{51.31578947368422}{7}{6};
    \dotswapgate{51.31578947368422}{5}{4};
    \dotswapgate{51.31578947368422}{3}{2};
    \dotswapgate{52.29323308270676}{6}{5};
    \dotswapgate{52.29323308270676}{4}{3};
    \dotswapgate{53.27067669172936}{7}{6};
    \dotswapgate{53.27067669172936}{5}{4};
    \dotswapgate{53.27067669172936}{3}{2};
    \wirelabel{\footnotesize 7}{54.492481203007515,7};
    \wirelabel{\footnotesize 4}{54.492481203007515,6};
    \wirelabel{\footnotesize 1}{54.492481203007515,5};
    \wirelabel{\footnotesize 3}{54.492481203007515,4};
    \wirelabel{\footnotesize 0}{54.492481203007515,3};
    \wirelabel{\footnotesize 2}{54.492481203007515,2};
    \wirelabel{\footnotesize 5}{54.492481203007515,1};
    \wirelabel{\footnotesize 6}{54.492481203007515,0};
    \biggate{$Q$}{55.95864661654139,4}{55.95864661654139,7};
    \biggate{$Q$}{55.95864661654139,0}{55.95864661654139,3};
    \dotswapgate{57.180451127819595}{6}{5};
    \dotswapgate{57.180451127819595}{3}{2};
    \dotswapgate{57.180451127819595}{1}{0};
    \dotswapgate{58.15789473684214}{5}{4};
    \dotswapgate{58.15789473684214}{2}{1};
    \dotswapgate{59.13533834586468}{6}{5};
    \dotswapgate{59.13533834586468}{4}{3};
    \dotswapgate{60.112781954887225}{7}{6};
    \dotswapgate{60.112781954887225}{5}{4};
    \dotswapgate{60.112781954887225}{3}{2};
    \dotswapgate{61.09022556390977}{6}{5};
    \dotswapgate{61.09022556390977}{4}{3};
    \dotswapgate{61.09022556390977}{2}{1};
    \dotswapgate{62.06766917293237}{7}{6};
    \dotswapgate{62.06766917293237}{5}{4};
    \dotswapgate{62.06766917293237}{3}{2};
    \dotswapgate{63.04511278195491}{4}{3};
    \wirelabel{\footnotesize 2}{64.26691729323312,7};
    \wirelabel{\footnotesize 3}{64.26691729323312,6};
    \wirelabel{\footnotesize 6}{64.26691729323312,5};
    \wirelabel{\footnotesize 0}{64.26691729323312,4};
    \wirelabel{\footnotesize 7}{64.26691729323312,3};
    \wirelabel{\footnotesize 1}{64.26691729323312,2};
    \wirelabel{\footnotesize 4}{64.26691729323312,1};
    \wirelabel{\footnotesize 5}{64.26691729323312,0};
  \end{qcircuit}
}
\]
\[
  \scalebox{0.45}{
  \begin{qcircuit}[scale=0.4]
    \grid{65.0}{0,1,2,3,4,5,6,7}
    \wirelabel{\footnotesize 2}{0.7558139534883708,7};
    \wirelabel{\footnotesize 3}{0.7558139534883708,6};
    \wirelabel{\footnotesize 6}{0.7558139534883708,5};
    \wirelabel{\footnotesize 0}{0.7558139534883708,4};
    \wirelabel{\footnotesize 7}{0.7558139534883708,3};
    \wirelabel{\footnotesize 1}{0.7558139534883708,2};
    \wirelabel{\footnotesize 4}{0.7558139534883708,1};
    \wirelabel{\footnotesize 5}{0.7558139534883708,0};
    \biggate{$Q$}{2.2674418604651123,4}{2.2674418604651123,7};
    \biggate{$Q$}{2.2674418604651123,0}{2.2674418604651123,3};
    \dotswapgate{3.5271317829457303}{6}{5};
    \dotswapgate{3.5271317829457303}{4}{3};
    \dotswapgate{4.534883720930225}{7}{6};
    \dotswapgate{4.534883720930225}{5}{4};
    \dotswapgate{4.534883720930225}{3}{2};
    \dotswapgate{5.542635658914719}{6}{5};
    \dotswapgate{5.542635658914719}{4}{3};
    \dotswapgate{6.550387596899213}{7}{6};
    \dotswapgate{6.550387596899213}{5}{4};
    \dotswapgate{6.550387596899213}{3}{2};
    \wirelabel{\footnotesize 7}{7.810077519379831,7};
    \wirelabel{\footnotesize 6}{7.810077519379831,6};
    \wirelabel{\footnotesize 1}{7.810077519379831,5};
    \wirelabel{\footnotesize 2}{7.810077519379831,4};
    \wirelabel{\footnotesize 0}{7.810077519379831,3};
    \wirelabel{\footnotesize 3}{7.810077519379831,2};
    \wirelabel{\footnotesize 4}{7.810077519379831,1};
    \wirelabel{\footnotesize 5}{7.810077519379831,0};
    \biggate{$Q$}{9.321705426356573,4}{9.321705426356573,7};
    \biggate{$Q$}{9.321705426356573,0}{9.321705426356573,3};
    \dotswapgate{10.58139534883719}{6}{5};
    \dotswapgate{10.58139534883719}{2}{1};
    \dotswapgate{11.589147286821685}{5}{4};
    \dotswapgate{11.589147286821685}{3}{2};
    \dotswapgate{11.589147286821685}{1}{0};
    \dotswapgate{12.59689922480618}{6}{5};
    \dotswapgate{12.59689922480618}{4}{3};
    \dotswapgate{12.59689922480618}{2}{1};
    \dotswapgate{13.604651162790674}{7}{6};
    \dotswapgate{13.604651162790674}{5}{4};
    \dotswapgate{13.604651162790674}{3}{2};
    \dotswapgate{14.612403100775168}{6}{5};
    \dotswapgate{14.612403100775168}{4}{3};
    \dotswapgate{14.612403100775168}{2}{1};
    \dotswapgate{15.620155038759663}{5}{4};
    \dotswapgate{15.620155038759663}{3}{2};
    \dotswapgate{15.620155038759663}{1}{0};
    \dotswapgate{16.627906976744157}{6}{5};
    \dotswapgate{16.627906976744157}{4}{3};
    \wirelabel{\footnotesize 2}{17.88759689922483,7};
    \wirelabel{\footnotesize 5}{17.88759689922483,6};
    \wirelabel{\footnotesize 4}{17.88759689922483,5};
    \wirelabel{\footnotesize 0}{17.88759689922483,4};
    \wirelabel{\footnotesize 7}{17.88759689922483,3};
    \wirelabel{\footnotesize 1}{17.88759689922483,2};
    \wirelabel{\footnotesize 3}{17.88759689922483,1};
    \wirelabel{\footnotesize 6}{17.88759689922483,0};
    \biggate{$Q$}{19.399224806201573,4}{19.399224806201573,7};
    \biggate{$Q$}{19.399224806201573,0}{19.399224806201573,3};
    \dotswapgate{20.65891472868219}{6}{5};
    \dotswapgate{20.65891472868219}{4}{3};
    \dotswapgate{21.666666666666686}{7}{6};
    \dotswapgate{21.666666666666686}{5}{4};
    \dotswapgate{21.666666666666686}{3}{2};
    \dotswapgate{22.67441860465118}{6}{5};
    \dotswapgate{22.67441860465118}{4}{3};
    \dotswapgate{23.682170542635674}{7}{6};
    \dotswapgate{23.682170542635674}{5}{4};
    \dotswapgate{23.682170542635674}{3}{2};
    \wirelabel{\footnotesize 7}{24.941860465116292,7};
    \wirelabel{\footnotesize 4}{24.941860465116292,6};
    \wirelabel{\footnotesize 1}{24.941860465116292,5};
    \wirelabel{\footnotesize 2}{24.941860465116292,4};
    \wirelabel{\footnotesize 0}{24.941860465116292,3};
    \wirelabel{\footnotesize 5}{24.941860465116292,2};
    \wirelabel{\footnotesize 3}{24.941860465116292,1};
    \wirelabel{\footnotesize 6}{24.941860465116292,0};
    \biggate{$Q$}{26.453488372093034,4}{26.453488372093034,7};
    \biggate{$Q$}{26.453488372093034,0}{26.453488372093034,3};
    \dotswapgate{27.713178294573652}{7}{6};
    \dotswapgate{27.713178294573652}{4}{3};
    \dotswapgate{27.713178294573652}{2}{1};
    \dotswapgate{28.720930232558146}{6}{5};
    \dotswapgate{28.720930232558146}{3}{2};
    \dotswapgate{29.72868217054264}{7}{6};
    \dotswapgate{29.72868217054264}{5}{4};
    \dotswapgate{30.736434108527135}{6}{5};
    \dotswapgate{30.736434108527135}{4}{3};
    \dotswapgate{31.74418604651163}{7}{6};
    \dotswapgate{31.74418604651163}{5}{4};
    \dotswapgate{31.74418604651163}{3}{2};
    \wirelabel{\footnotesize 0}{33.00387596899225,7};
    \wirelabel{\footnotesize 1}{33.00387596899225,6};
    \wirelabel{\footnotesize 3}{33.00387596899225,5};
    \wirelabel{\footnotesize 4}{33.00387596899225,4};
    \wirelabel{\footnotesize 2}{33.00387596899225,3};
    \wirelabel{\footnotesize 7}{33.00387596899225,2};
    \wirelabel{\footnotesize 5}{33.00387596899225,1};
    \wirelabel{\footnotesize 6}{33.00387596899225,0};
    \biggate{$Q$}{34.51550387596899,4}{34.51550387596899,7};
    \biggate{$Q$}{34.51550387596899,0}{34.51550387596899,3};
    \dotswapgate{35.77519379844961}{5}{4};
    \dotswapgate{35.77519379844961}{1}{0};
    \dotswapgate{36.7829457364341}{4}{3};
    \dotswapgate{36.7829457364341}{2}{1};
    \dotswapgate{37.790697674418595}{3}{2};
    \dotswapgate{38.79844961240309}{4}{3};
    \dotswapgate{38.79844961240309}{2}{1};
    \wirelabel{\footnotesize 0}{40.05813953488371,7};
    \wirelabel{\footnotesize 1}{40.05813953488371,6};
    \wirelabel{\footnotesize 4}{40.05813953488371,5};
    \wirelabel{\footnotesize 6}{40.05813953488371,4};
    \wirelabel{\footnotesize 2}{40.05813953488371,3};
    \wirelabel{\footnotesize 7}{40.05813953488371,2};
    \wirelabel{\footnotesize 3}{40.05813953488371,1};
    \wirelabel{\footnotesize 5}{40.05813953488371,0};
    \biggate{$Q$}{41.56976744186045,4}{41.56976744186045,7};
    \biggate{$Q$}{41.56976744186045,0}{41.56976744186045,3};
    \dotswapgate{42.82945736434107}{6}{5};
    \dotswapgate{42.82945736434107}{4}{3};
    \dotswapgate{43.83720930232556}{7}{6};
    \dotswapgate{43.83720930232556}{5}{4};
    \dotswapgate{43.83720930232556}{3}{2};
    \dotswapgate{44.844961240310056}{6}{5};
    \dotswapgate{44.844961240310056}{4}{3};
    \dotswapgate{45.85271317829455}{7}{6};
    \dotswapgate{45.85271317829455}{5}{4};
    \dotswapgate{45.85271317829455}{3}{2};
    \wirelabel{\footnotesize 2}{47.11240310077517,7};
    \wirelabel{\footnotesize 4}{47.11240310077517,6};
    \wirelabel{\footnotesize 7}{47.11240310077517,5};
    \wirelabel{\footnotesize 0}{47.11240310077517,4};
    \wirelabel{\footnotesize 6}{47.11240310077517,3};
    \wirelabel{\footnotesize 1}{47.11240310077517,2};
    \wirelabel{\footnotesize 3}{47.11240310077517,1};
    \wirelabel{\footnotesize 5}{47.11240310077517,0};
    \biggate{$Q$}{48.62403100775191,4}{48.62403100775191,7};
    \biggate{$Q$}{48.62403100775191,0}{48.62403100775191,3};
    \dotswapgate{49.883720930232585}{6}{5};
    \dotswapgate{49.883720930232585}{3}{2};
    \dotswapgate{50.89147286821708}{5}{4};
    \dotswapgate{51.89922480620157}{6}{5};
    \dotswapgate{51.89922480620157}{4}{3};
    \dotswapgate{52.90697674418607}{7}{6};
    \dotswapgate{52.90697674418607}{5}{4};
    \dotswapgate{52.90697674418607}{3}{2};
    \dotswapgate{53.91472868217056}{6}{5};
    \dotswapgate{53.91472868217056}{4}{3};
    \dotswapgate{53.91472868217056}{2}{1};
    \dotswapgate{54.922480620155056}{5}{4};
    \dotswapgate{54.922480620155056}{3}{2};
    \dotswapgate{55.93023255813955}{4}{3};
    \wirelabel{\footnotesize 0}{57.18992248062017,7};
    \wirelabel{\footnotesize 1}{57.18992248062017,6};
    \wirelabel{\footnotesize 6}{57.18992248062017,5};
    \wirelabel{\footnotesize 3}{57.18992248062017,4};
    \wirelabel{\footnotesize 2}{57.18992248062017,3};
    \wirelabel{\footnotesize 7}{57.18992248062017,2};
    \wirelabel{\footnotesize 4}{57.18992248062017,1};
    \wirelabel{\footnotesize 5}{57.18992248062017,0};
    \biggate{$Q$}{58.70155038759691,4}{58.70155038759691,7};
    \biggate{$Q$}{58.70155038759691,0}{58.70155038759691,3};
    \dotswapgate{59.96124031007753}{7}{6};
    \dotswapgate{59.96124031007753}{5}{4};
    \dotswapgate{59.96124031007753}{3}{2};
    \dotswapgate{60.96899224806202}{6}{5};
    \dotswapgate{60.96899224806202}{4}{3};
    \dotswapgate{61.97674418604652}{7}{6};
    \dotswapgate{61.97674418604652}{5}{4};
    \dotswapgate{61.97674418604652}{3}{2};
    \dotswapgate{62.98449612403101}{6}{5};
    \dotswapgate{62.98449612403101}{4}{3};
    \wirelabel{\footnotesize 3}{64.24418604651163,7};
    \wirelabel{\footnotesize 7}{64.24418604651163,6};
    \wirelabel{\footnotesize 1}{64.24418604651163,5};
    \wirelabel{\footnotesize 2}{64.24418604651163,4};
    \wirelabel{\footnotesize 0}{64.24418604651163,3};
    \wirelabel{\footnotesize 6}{64.24418604651163,2};
    \wirelabel{\footnotesize 4}{64.24418604651163,1};
    \wirelabel{\footnotesize 5}{64.24418604651163,0};
  \end{qcircuit}
}
\]
\[
  \scalebox{0.45}{
  \begin{qcircuit}[scale=0.4]
    \grid{65.0}{0,1,2,3,4,5,6,7}
    \wirelabel{\footnotesize 3}{0.7330827067669361,7};
    \wirelabel{\footnotesize 7}{0.7330827067669361,6};
    \wirelabel{\footnotesize 1}{0.7330827067669361,5};
    \wirelabel{\footnotesize 2}{0.7330827067669361,4};
    \wirelabel{\footnotesize 0}{0.7330827067669361,3};
    \wirelabel{\footnotesize 6}{0.7330827067669361,2};
    \wirelabel{\footnotesize 4}{0.7330827067669361,1};
    \wirelabel{\footnotesize 5}{0.7330827067669361,0};
    \biggate{$Q$}{2.1992481203007515,4}{2.1992481203007515,7};
    \biggate{$Q$}{2.1992481203007515,0}{2.1992481203007515,3};
    \dotswapgate{3.4210526315789593}{6}{5};
    \dotswapgate{3.4210526315789593}{4}{3};
    \dotswapgate{3.4210526315789593}{2}{1};
    \dotswapgate{4.398496240601503}{7}{6};
    \dotswapgate{4.398496240601503}{5}{4};
    \dotswapgate{4.398496240601503}{1}{0};
    \dotswapgate{5.3759398496240465}{6}{5};
    \dotswapgate{5.3759398496240465}{4}{3};
    \dotswapgate{6.353383458646647}{7}{6};
    \dotswapgate{6.353383458646647}{5}{4};
    \dotswapgate{6.353383458646647}{3}{2};
    \dotswapgate{7.3308270676691905}{6}{5};
    \dotswapgate{7.3308270676691905}{4}{3};
    \wirelabel{\footnotesize 0}{8.552631578947398,7};
    \wirelabel{\footnotesize 2}{8.552631578947398,6};
    \wirelabel{\footnotesize 1}{8.552631578947398,5};
    \wirelabel{\footnotesize 4}{8.552631578947398,4};
    \wirelabel{\footnotesize 3}{8.552631578947398,3};
    \wirelabel{\footnotesize 7}{8.552631578947398,2};
    \wirelabel{\footnotesize 5}{8.552631578947398,1};
    \wirelabel{\footnotesize 6}{8.552631578947398,0};
    \biggate{$Q$}{10.018796992481214,4}{10.018796992481214,7};
    \biggate{$Q$}{10.018796992481214,0}{10.018796992481214,3};
    \dotswapgate{11.240601503759422}{7}{6};
    \dotswapgate{11.240601503759422}{5}{4};
    \dotswapgate{11.240601503759422}{3}{2};
    \dotswapgate{12.218045112781965}{6}{5};
    \dotswapgate{12.218045112781965}{4}{3};
    \dotswapgate{13.195488721804509}{7}{6};
    \dotswapgate{13.195488721804509}{5}{4};
    \dotswapgate{13.195488721804509}{3}{2};
    \dotswapgate{14.172932330827052}{6}{5};
    \dotswapgate{14.172932330827052}{4}{3};
    \wirelabel{\footnotesize 4}{15.39473684210526,7};
    \wirelabel{\footnotesize 7}{15.39473684210526,6};
    \wirelabel{\footnotesize 2}{15.39473684210526,5};
    \wirelabel{\footnotesize 3}{15.39473684210526,4};
    \wirelabel{\footnotesize 0}{15.39473684210526,3};
    \wirelabel{\footnotesize 1}{15.39473684210526,2};
    \wirelabel{\footnotesize 5}{15.39473684210526,1};
    \wirelabel{\footnotesize 6}{15.39473684210526,0};
    \biggate{$Q$}{16.860902255639132,4}{16.860902255639132,7};
    \biggate{$Q$}{16.860902255639132,0}{16.860902255639132,3};
    \dotswapgate{18.082706766917283}{7}{6};
    \dotswapgate{18.082706766917283}{4}{3};
    \dotswapgate{19.060150375939884}{6}{5};
    \dotswapgate{19.060150375939884}{3}{2};
    \dotswapgate{20.037593984962427}{7}{6};
    \dotswapgate{20.037593984962427}{5}{4};
    \dotswapgate{20.037593984962427}{2}{1};
    \dotswapgate{21.01503759398497}{6}{5};
    \dotswapgate{21.01503759398497}{4}{3};
    \dotswapgate{21.992481203007515}{7}{6};
    \dotswapgate{21.992481203007515}{5}{4};
    \dotswapgate{21.992481203007515}{3}{2};
    \dotswapgate{22.969924812030058}{6}{5};
    \dotswapgate{22.969924812030058}{4}{3};
    \dotswapgate{22.969924812030058}{2}{1};
    \dotswapgate{23.94736842105266}{3}{2};
    \wirelabel{\footnotesize 0}{25.16917293233081,7};
    \wirelabel{\footnotesize 1}{25.16917293233081,6};
    \wirelabel{\footnotesize 2}{25.16917293233081,5};
    \wirelabel{\footnotesize 5}{25.16917293233081,4};
    \wirelabel{\footnotesize 3}{25.16917293233081,3};
    \wirelabel{\footnotesize 7}{25.16917293233081,2};
    \wirelabel{\footnotesize 4}{25.16917293233081,1};
    \wirelabel{\footnotesize 6}{25.16917293233081,0};
    \biggate{$Q$}{26.63533834586468,4}{26.63533834586468,7};
    \biggate{$Q$}{26.63533834586468,0}{26.63533834586468,3};
    \dotswapgate{27.85714285714289}{7}{6};
    \dotswapgate{27.85714285714289}{5}{4};
    \dotswapgate{27.85714285714289}{3}{2};
    \dotswapgate{28.834586466165433}{6}{5};
    \dotswapgate{28.834586466165433}{4}{3};
    \dotswapgate{29.812030075187977}{7}{6};
    \dotswapgate{29.812030075187977}{5}{4};
    \dotswapgate{29.812030075187977}{3}{2};
    \dotswapgate{30.78947368421052}{6}{5};
    \dotswapgate{30.78947368421052}{4}{3};
    \wirelabel{\footnotesize 5}{32.01127819548873,7};
    \wirelabel{\footnotesize 7}{32.01127819548873,6};
    \wirelabel{\footnotesize 1}{32.01127819548873,5};
    \wirelabel{\footnotesize 3}{32.01127819548873,4};
    \wirelabel{\footnotesize 0}{32.01127819548873,3};
    \wirelabel{\footnotesize 2}{32.01127819548873,2};
    \wirelabel{\footnotesize 4}{32.01127819548873,1};
    \wirelabel{\footnotesize 6}{32.01127819548873,0};
    \biggate{$Q$}{33.477443609022544,4}{33.477443609022544,7};
    \biggate{$Q$}{33.477443609022544,0}{33.477443609022544,3};
    \dotswapgate{34.69924812030075}{6}{5};
    \dotswapgate{34.69924812030075}{4}{3};
    \dotswapgate{34.69924812030075}{2}{1};
    \dotswapgate{35.676691729323295}{5}{4};
    \dotswapgate{35.676691729323295}{3}{2};
    \dotswapgate{36.654135338345895}{4}{3};
    \dotswapgate{36.654135338345895}{2}{1};
    \dotswapgate{37.63157894736844}{5}{4};
    \dotswapgate{37.63157894736844}{3}{2};
    \dotswapgate{37.63157894736844}{1}{0};
    \dotswapgate{38.60902255639098}{6}{5};
    \dotswapgate{38.60902255639098}{4}{3};
    \dotswapgate{38.60902255639098}{2}{1};
    \dotswapgate{39.586466165413526}{7}{6};
    \dotswapgate{39.586466165413526}{1}{0};
    \wirelabel{\footnotesize 4}{40.808270676691734,7};
    \wirelabel{\footnotesize 5}{40.808270676691734,6};
    \wirelabel{\footnotesize 1}{40.808270676691734,5};
    \wirelabel{\footnotesize 2}{40.808270676691734,4};
    \wirelabel{\footnotesize 0}{40.808270676691734,3};
    \wirelabel{\footnotesize 6}{40.808270676691734,2};
    \wirelabel{\footnotesize 3}{40.808270676691734,1};
    \wirelabel{\footnotesize 7}{40.808270676691734,0};
    \biggate{$Q$}{42.27443609022555,4}{42.27443609022555,7};
    \biggate{$Q$}{42.27443609022555,0}{42.27443609022555,3};
    \dotswapgate{43.49624060150376}{7}{6};
    \dotswapgate{43.49624060150376}{5}{4};
    \dotswapgate{43.49624060150376}{3}{2};
    \dotswapgate{44.4736842105263}{6}{5};
    \dotswapgate{44.4736842105263}{4}{3};
    \dotswapgate{44.4736842105263}{2}{1};
    \dotswapgate{45.4511278195489}{5}{4};
    \dotswapgate{45.4511278195489}{3}{2};
    \dotswapgate{46.428571428571445}{6}{5};
    \dotswapgate{46.428571428571445}{4}{3};
    \dotswapgate{46.428571428571445}{2}{1};
    \dotswapgate{47.40601503759399}{3}{2};
    \dotswapgate{48.38345864661653}{2}{1};
    \wirelabel{\footnotesize 5}{49.60526315789474,7};
    \wirelabel{\footnotesize 6}{49.60526315789474,6};
    \wirelabel{\footnotesize 2}{49.60526315789474,5};
    \wirelabel{\footnotesize 3}{49.60526315789474,4};
    \wirelabel{\footnotesize 0}{49.60526315789474,3};
    \wirelabel{\footnotesize 1}{49.60526315789474,2};
    \wirelabel{\footnotesize 4}{49.60526315789474,1};
    \wirelabel{\footnotesize 7}{49.60526315789474,0};
    \biggate{$Q$}{51.071428571428555,4}{51.071428571428555,7};
    \biggate{$Q$}{51.071428571428555,0}{51.071428571428555,3};
    \dotswapgate{52.29323308270676}{6}{5};
    \dotswapgate{52.29323308270676}{4}{3};
    \dotswapgate{53.27067669172931}{5}{4};
    \dotswapgate{53.27067669172931}{3}{2};
    \dotswapgate{54.24812030075191}{6}{5};
    \dotswapgate{54.24812030075191}{4}{3};
    \dotswapgate{55.22556390977445}{7}{6};
    \dotswapgate{55.22556390977445}{5}{4};
    \dotswapgate{55.22556390977445}{3}{2};
    \dotswapgate{56.203007518796994}{6}{5};
    \dotswapgate{56.203007518796994}{4}{3};
    \wirelabel{\footnotesize 0}{57.4248120300752,7};
    \wirelabel{\footnotesize 1}{57.4248120300752,6};
    \wirelabel{\footnotesize 5}{57.4248120300752,5};
    \wirelabel{\footnotesize 3}{57.4248120300752,4};
    \wirelabel{\footnotesize 2}{57.4248120300752,3};
    \wirelabel{\footnotesize 6}{57.4248120300752,2};
    \wirelabel{\footnotesize 4}{57.4248120300752,1};
    \wirelabel{\footnotesize 7}{57.4248120300752,0};
    \biggate{$Q$}{58.89097744360902,4}{58.89097744360902,7};
    \biggate{$Q$}{58.89097744360902,0}{58.89097744360902,3};
    \dotswapgate{60.112781954887225}{5}{4};
    \dotswapgate{60.112781954887225}{3}{2};
    \dotswapgate{61.09022556390977}{4}{3};
    \dotswapgate{62.06766917293231}{5}{4};
    \dotswapgate{62.06766917293231}{3}{2};
    \dotswapgate{63.04511278195491}{4}{3};
    \wirelabel{\footnotesize 0}{64.26691729323306,7};
    \wirelabel{\footnotesize 1}{64.26691729323306,6};
    \wirelabel{\footnotesize 6}{64.26691729323306,5};
    \wirelabel{\footnotesize 2}{64.26691729323306,4};
    \wirelabel{\footnotesize 3}{64.26691729323306,3};
    \wirelabel{\footnotesize 5}{64.26691729323306,2};
    \wirelabel{\footnotesize 4}{64.26691729323306,1};
    \wirelabel{\footnotesize 7}{64.26691729323306,0};
  \end{qcircuit}
}
\]
\[
  \scalebox{0.45}{
  \begin{qcircuit}[scale=0.4]
    \grid{65.0}{0,1,2,3,4,5,6,7}
    \wirelabel{\footnotesize 0}{0.8478260869566157,7};
    \wirelabel{\footnotesize 1}{0.8478260869566157,6};
    \wirelabel{\footnotesize 6}{0.8478260869566157,5};
    \wirelabel{\footnotesize 2}{0.8478260869566157,4};
    \wirelabel{\footnotesize 3}{0.8478260869566157,3};
    \wirelabel{\footnotesize 5}{0.8478260869566157,2};
    \wirelabel{\footnotesize 4}{0.8478260869566157,1};
    \wirelabel{\footnotesize 7}{0.8478260869566157,0};
    \biggate{$Q$}{2.5434782608696196,4}{2.5434782608696196,7};
    \biggate{$Q$}{2.5434782608696196,0}{2.5434782608696196,3};
    \dotswapgate{3.956521739130494}{5}{4};
    \dotswapgate{3.956521739130494}{2}{1};
    \dotswapgate{5.0869565217391255}{6}{5};
    \dotswapgate{5.0869565217391255}{4}{3};
    \dotswapgate{6.217391304347871}{3}{2};
    \wirelabel{\footnotesize 0}{7.630434782608745,7};
    \wirelabel{\footnotesize 2}{7.630434782608745,6};
    \wirelabel{\footnotesize 1}{7.630434782608745,5};
    \wirelabel{\footnotesize 3}{7.630434782608745,4};
    \wirelabel{\footnotesize 4}{7.630434782608745,3};
    \wirelabel{\footnotesize 6}{7.630434782608745,2};
    \wirelabel{\footnotesize 5}{7.630434782608745,1};
    \wirelabel{\footnotesize 7}{7.630434782608745,0};
    \biggate{$Q$}{9.326086956521749,4}{9.326086956521749,7};
    \biggate{$Q$}{9.326086956521749,0}{9.326086956521749,3};
    \dotswapgate{10.739130434782624}{7}{6};
    \dotswapgate{10.739130434782624}{4}{3};
    \dotswapgate{11.869565217391369}{6}{5};
    \dotswapgate{11.869565217391369}{3}{2};
    \dotswapgate{13.0}{7}{6};
    \dotswapgate{13.0}{5}{4};
    \dotswapgate{14.130434782608745}{6}{5};
    \dotswapgate{14.130434782608745}{4}{3};
    \dotswapgate{15.26086956521749}{7}{6};
    \dotswapgate{15.26086956521749}{5}{4};
    \dotswapgate{16.39130434782612}{6}{5};
    \dotswapgate{17.521739130434867}{7}{6};
    \wirelabel{\footnotesize 6}{18.93478260869574,7};
    \wirelabel{\footnotesize 4}{18.93478260869574,6};
    \wirelabel{\footnotesize 1}{18.93478260869574,5};
    \wirelabel{\footnotesize 2}{18.93478260869574,4};
    \wirelabel{\footnotesize 0}{18.93478260869574,3};
    \wirelabel{\footnotesize 3}{18.93478260869574,2};
    \wirelabel{\footnotesize 5}{18.93478260869574,1};
    \wirelabel{\footnotesize 7}{18.93478260869574,0};
    \biggate{$Q$}{20.630434782608745,4}{20.630434782608745,7};
    \biggate{$Q$}{20.630434782608745,0}{20.630434782608745,3};
    \dotswapgate{22.04347826086962}{7}{6};
    \dotswapgate{22.04347826086962}{5}{4};
    \dotswapgate{22.04347826086962}{3}{2};
    \dotswapgate{23.173913043478365}{4}{3};
    \dotswapgate{24.304347826086996}{5}{4};
    \dotswapgate{24.304347826086996}{3}{2};
    \dotswapgate{25.43478260869574}{6}{5};
    \dotswapgate{25.43478260869574}{4}{3};
    \dotswapgate{26.565217391304373}{7}{6};
    \wirelabel{\footnotesize 3}{27.978260869565247,7};
    \wirelabel{\footnotesize 4}{27.978260869565247,6};
    \wirelabel{\footnotesize 6}{27.978260869565247,5};
    \wirelabel{\footnotesize 0}{27.978260869565247,4};
    \wirelabel{\footnotesize 2}{27.978260869565247,3};
    \wirelabel{\footnotesize 1}{27.978260869565247,2};
    \wirelabel{\footnotesize 5}{27.978260869565247,1};
    \wirelabel{\footnotesize 7}{27.978260869565247,0};
    \biggate{$Q$}{29.673913043478365,4}{29.673913043478365,7};
    \biggate{$Q$}{29.673913043478365,0}{29.673913043478365,3};
    \dotswapgate{31.086956521739125}{5}{4};
    \dotswapgate{31.086956521739125}{2}{1};
    \dotswapgate{32.21739130434787}{6}{5};
    \dotswapgate{32.21739130434787}{4}{3};
    \dotswapgate{33.347826086956616}{5}{4};
    \dotswapgate{33.347826086956616}{3}{2};
    \dotswapgate{34.47826086956525}{6}{5};
    \dotswapgate{34.47826086956525}{4}{3};
    \dotswapgate{34.47826086956525}{2}{1};
    \dotswapgate{35.60869565217399}{5}{4};
    \wirelabel{\footnotesize 3}{37.02173913043487,7};
    \wirelabel{\footnotesize 2}{37.02173913043487,6};
    \wirelabel{\footnotesize 5}{37.02173913043487,5};
    \wirelabel{\footnotesize 0}{37.02173913043487,4};
    \wirelabel{\footnotesize 4}{37.02173913043487,3};
    \wirelabel{\footnotesize 1}{37.02173913043487,2};
    \wirelabel{\footnotesize 6}{37.02173913043487,1};
    \wirelabel{\footnotesize 7}{37.02173913043487,0};
    \biggate{$Q$}{38.71739130434787,4}{38.71739130434787,7};
    \biggate{$Q$}{38.71739130434787,0}{38.71739130434787,3};
    \dotswapgate{40.130434782608745}{7}{6};
    \dotswapgate{40.130434782608745}{5}{4};
    \dotswapgate{40.130434782608745}{3}{2};
    \dotswapgate{41.26086956521749}{6}{5};
    \dotswapgate{41.26086956521749}{4}{3};
    \dotswapgate{42.39130434782612}{7}{6};
    \dotswapgate{42.39130434782612}{5}{4};
    \dotswapgate{42.39130434782612}{3}{2};
    \dotswapgate{43.52173913043487}{6}{5};
    \dotswapgate{43.52173913043487}{4}{3};
    \dotswapgate{44.6521739130435}{5}{4};
    \dotswapgate{44.6521739130435}{3}{2};
    \dotswapgate{45.78260869565224}{4}{3};
    \wirelabel{\footnotesize 0}{47.19565217391312,7};
    \wirelabel{\footnotesize 1}{47.19565217391312,6};
    \wirelabel{\footnotesize 4}{47.19565217391312,5};
    \wirelabel{\footnotesize 5}{47.19565217391312,4};
    \wirelabel{\footnotesize 2}{47.19565217391312,3};
    \wirelabel{\footnotesize 3}{47.19565217391312,2};
    \wirelabel{\footnotesize 6}{47.19565217391312,1};
    \wirelabel{\footnotesize 7}{47.19565217391312,0};
    \biggate{$Q$}{48.89130434782612,4}{48.89130434782612,7};
    \biggate{$Q$}{48.89130434782612,0}{48.89130434782612,3};
    \dotswapgate{50.304347826086996}{7}{6};
    \dotswapgate{50.304347826086996}{4}{3};
    \dotswapgate{51.43478260869574}{6}{5};
    \dotswapgate{51.43478260869574}{3}{2};
    \dotswapgate{52.56521739130437}{7}{6};
    \dotswapgate{52.56521739130437}{5}{4};
    \dotswapgate{53.69565217391312}{6}{5};
    \dotswapgate{53.69565217391312}{4}{3};
    \dotswapgate{54.82608695652175}{7}{6};
    \wirelabel{\footnotesize 2}{56.23913043478262,7};
    \wirelabel{\footnotesize 4}{56.23913043478262,6};
    \wirelabel{\footnotesize 1}{56.23913043478262,5};
    \wirelabel{\footnotesize 3}{56.23913043478262,4};
    \wirelabel{\footnotesize 0}{56.23913043478262,3};
    \wirelabel{\footnotesize 5}{56.23913043478262,2};
    \wirelabel{\footnotesize 6}{56.23913043478262,1};
    \wirelabel{\footnotesize 7}{56.23913043478262,0};
    \biggate{$Q$}{57.93478260869574,4}{57.93478260869574,7};
    \biggate{$Q$}{57.93478260869574,0}{57.93478260869574,3};
    \dotswapgate{59.347826086956616}{6}{5};
    \dotswapgate{59.347826086956616}{4}{3};
    \dotswapgate{60.47826086956525}{7}{6};
    \dotswapgate{60.47826086956525}{5}{4};
    \dotswapgate{61.60869565217399}{6}{5};
    \dotswapgate{61.60869565217399}{4}{3};
    \dotswapgate{62.73913043478262}{7}{6};
    \wirelabel{\footnotesize 0}{64.1521739130435,7};
    \wirelabel{\footnotesize 1}{64.1521739130435,6};
    \wirelabel{\footnotesize 2}{64.1521739130435,5};
    \wirelabel{\footnotesize 3}{64.1521739130435,4};
    \wirelabel{\footnotesize 4}{64.1521739130435,3};
    \wirelabel{\footnotesize 5}{64.1521739130435,2};
    \wirelabel{\footnotesize 6}{64.1521739130435,1};
    \wirelabel{\footnotesize 7}{64.1521739130435,0};
  \end{qcircuit}
}
\]
\caption{An example ski-lift schedule for 8 orbitals. This circuit
  should be read from top left to bottom right like a musical
  score. Singletons, pairs, triples, and quads are denotes $S$, $P$,
  $T$, and $Q$, respectively. Note that as the number of orbitals
  increases, quad stages will dominate the schedule. Also note that
  the transition from one quad stage to the next typically has a
  permutation depth of 4, but occasionally has a greater permutation
  depth. These greater-depth cases happen only $O(1/m)$ of the time,
  so the average permutation depth per stage is asymptotically
  constant.}
\label{fig-schedule}
\end{figure}

Figure~\ref{fig-stages} shows a singleton stage, pair stage, triple
stage, and quad stage in the case of 8 orbitals and 2 precision bits.
Figure~\ref{fig-schedule} shows a ski-lift schedule for a single
Trotter-Suzuki step for 8 orbitals, including the fermionic
permutations. We did not show any precision bits in this schedule;
each of the boxes marking singleton, pair, triple, and quad circuits
must be expanded along the lines of Figure~\ref{fig-stages}, and the
precision bits enter the picture at that point.

Comparing our implementation of a Trotter-Suzuki step for ground state
energy estimation with the baseline implementation of Whitfield et
al. {\cite{WBA2011}}, for $m=120$ orbitals and $b=1$ precision bit, we
find a decrease in circuit rotation depth (counting only rotation
gates) by a factor of 720. Of this, a factor of 8 is due to improved
Hamiltonian circuits (Section~\ref{ssec:hamilton}), and a factor of 90
is due to ski-lift parallelization and parallel controlled rotations
(Sections~\ref{ssec:parallel-controlled-rotations} and
{\ref{ssec:skilift}}). The circuit width increases by a factor of
approximately 2. This is due to the fact that our parallel scheduling
makes use of qubits that were previously idle.

In addition to these improvements in the rotation depth, we also
optimized the fermionic swap depth, requiring only a constant depth of
non-rotation gates per stage on average. By contrast, the baseline
implementation requires Clifford basis changes whose average depth per
stage is linear in the number of orbitals. Moreover, the baseline
circuit's length increases linearly with the number of precision bits
$b$ while its width remains essentially constant, whereas in our
implementation, the width increases linearly with $b$ while the length
stays constant (assuming that sets of gates that can be done in
constant time in lattice surgery are counted as constant depth).

% ......................................................................

% ......................................................................

\section*{Acknowledgements}

This research was supported under DARPA Research Contract
HR001122C0066. We would like to thank Ryan MacDonell and Yunseong Nam
for helpful comments. We would also like to thank Fedor Nazarov and
David Speyer for answering our question on MathOverflow.

% ----------------------------------------------------------------------
\newpage
\bibliographystyle{abbrv}
\bibliography{skilift}

\end{document}